\documentclass[10pt,twocolumn]{article}  
\usepackage[utf8]{inputenc}
\usepackage{graphicx}
\usepackage{txfonts}
\usepackage{float}
\usepackage{natbib}
\usepackage{aas_macros}
\usepackage[a4paper, total={7.3in, 8.3in}]{geometry}

\title{Nautilus multi-grain model: Importance of cosmic-ray-induced desorption in determining the chemical abundances in the ISM}
\author{Wasim Iqbal and Valentine Wakelam \\
            \\
        Laboratoire d'astrophysique de Bordeaux, Univ. Bordeaux, \\
        CNRS, B18N, all\'ee Geoffroy Saint-Hilaire, 33615 Pessac, France\\
        \\
        email: wasimiqbal2009@gmail.com~~~~~~~\\
        email: valentine.wakelam@u-bordeaux.fr\\
        }
\begin{document} 

\date{2018-02-12}

%
%
%
%
%
%
%
%
%
%

\maketitle
\abstract
 {Species abundances in the interstellar medium (ISM) strongly depend on the chemistry occurring at the surfaces of the dust grains. To describe the complexity of the chemistry, various numerical models have been constructed. In most of these models, the grains are described by a single size of 0.1$\mu$m.}
%
%
 {We study the impact on the abundances of many species observed in the cold cores by considering several grain sizes in the Nautilus multi-grain model.}
%
%
 {We used grain sizes with radii in the range of $0.005\mu$m to $0.25\mu$m. We sampled this range in many bins. We used the previously published, MRN and WD grain size distributions to calculate the number density of grains in each bin. Other parameters such as the grain surface temperature or the cosmic-ray-induced desorption rates also vary with grain sizes.}
%
%
 {We present the abundances of various molecules in the gas phase and also on the dust surface at different time intervals during the simulation. We present a comparative study of results obtained using the single grain and the multi-grain models. We also compare our results with the observed abundances in TMC-1 and L134N clouds.}
%
%
 {We show that the grain size, the grain size dependent surface temperature and the peak surface temperature induced by cosmic ray collisions, play key roles in determining the ice and the gas phase abundances of various molecules. We also show that the differences between the MRN and the WD models are crucial for better fitting the observed abundances in different regions in the ISM. We show that the small grains play a very important role in the enrichment of the gas phase with the species which are mainly formed on the grain surface, as non-thermal desorption induced by collisions of cosmic ray particles is very efficient on the   small  grains.}
\section{Introduction}
\label{intro}
The interstellar medium (ISM) is rich in many molecular and atomic species either in the positively or the negatively charged states or in the neutral form. The ISM is also filled with different types of radiation fields such as UV photons (called interstellar radiation field) and cosmic rays. The local physical and irradiation conditions determine the main physio-chemical processes governing the chemical composition of the interstellar matter. In this context, the interstellar dust grains are of particular importance as they act as the surfaces where the gaseous species (atomic or molecular) can stick, diffuse, and react with each other to form new species. Numerical models based on the rate equation approximation are the most popular tool for studying the gas-grain chemistry and to simulate the evolution of complex chemical species with time, due to their very rapid execution \citep{Wakelam2013}. In general,  as a simplification, these models consider a single grain size with radius 0.1$\mu$m as a representative of all interstellar grains, which essentially results in all grains having the same physical and chemical properties (\cite{Hasegawa1992, Hasegawa1993b} and more recently \cite{Garrod2008,Wakelam08,Wakelam2010,Ruaud2016}). This assumption greatly reduces the difficulties in implementation of rate equations and also saves on computing time. There are a few published studies where the authors studied the impact of this assumption by considering several grain sizes in their model. \cite{Acharyya2011}, for instance, assumed five different grain sizes but the dust temperature was kept constant for all grain sizes. The conclusion of his work was that gas phase abundance of any species forming on the surface of dust grains depends on the effective total surface area of dust grains and this can also be achieved by using a single-grain model and proper number density of grains so that the total surface area remains constant. Following this work, \cite{Pauly2016} and \cite{Ge2016} developed models in which they used grains of different sizes. There were however two limitations to these works: They used a small number of grain sizes, and did not, as far as we know, consider the effect of cosmic-ray induced desorption as a function of grain radius, which strongly depends on the grain size. As \cite{Herbst2006} showed, the surface temperature of very small grains (radius $\approx$ 0.005$\mu$m) can rise above 300\ K, whereas that of big 0.25$\mu$m grains only reaches about 50~K. This can result in significant desorption of species from the surface of  small grains while big grains can remain very much unaffected.

In this work, we have revisited the effect of considering several grain sizes in our gas-grain model Nautilus in the conditions of cold cores, where the cosmic ray induced desorption can play an important role. While changing the size of the grains, several processes and model parameters are changed, such as the number density of each grain population, the surface temperature, and the peak surface temperature due to cosmic ray collisions with dust grains, among others. 

The paper is organized as follows. In Sect.~\ref{model}, we present the Nautilus gas-grain model and the modifications made to include the different grain sizes. In Sect.~\ref{results}, we present our results obtained using the multi-grain model and also compare these results with the single-grain Nautilus model. In Sect. \ref{sec:comp}, we compare our results with observations in TMC-1 and L134N clouds,  and in Sect. \ref{sec:discussion}, we discuss the effect of scaling the cosmic ray peak duration on   small  grains. Some conclusions are given in the last section. 

\section{Nautilus gas-grain code}
\label{model}
The Nautilus gas-grain code uses rate equation approximation (\cite{Hasegawa1992,Hasegawa1993b}) to simulate the chemical evolution in the ISM. The latest version of this code is based on the three-phase model from \cite{Ruaud2016}. It simulates the chemical evolution in the gas, on the grain surface, and within the grain mantle with all three phases coupled to each other. The gas-phase chemistry is based on the kida.uva.2014 public network \citep{Wakelam2015} while the surface chemistry is the same as in \cite {Ruaud2016}. In addition to the gas-phase chemistry, the model includes the physisorption of neutral species on the surface of grains, the diffusion of these species, and their reactions. Desorption of species can only occur from the surface (not the mantle) but the surface is of course reconstructed by the mantle species as the surface species evaporate. Similarly, as species are accreted on the surface, the species from the surface are incorporated to the mantle. In addition to thermal desorption, we consider nonthermal desorption processes such as cosmic-ray-induced desorption, UV (direct and indirect) photo-desorption, and chemical desorption. All details of the model are described in \cite{Ruaud2016}. 
\subsection{Implementation of multi-grain network and grain size distribution}
\label{sec:multi_grain}
\begin{figure*}
   \centering
   \includegraphics[width=.97\textwidth,trim = 0cm 0cm 0cm 0cm, clip,angle=0]{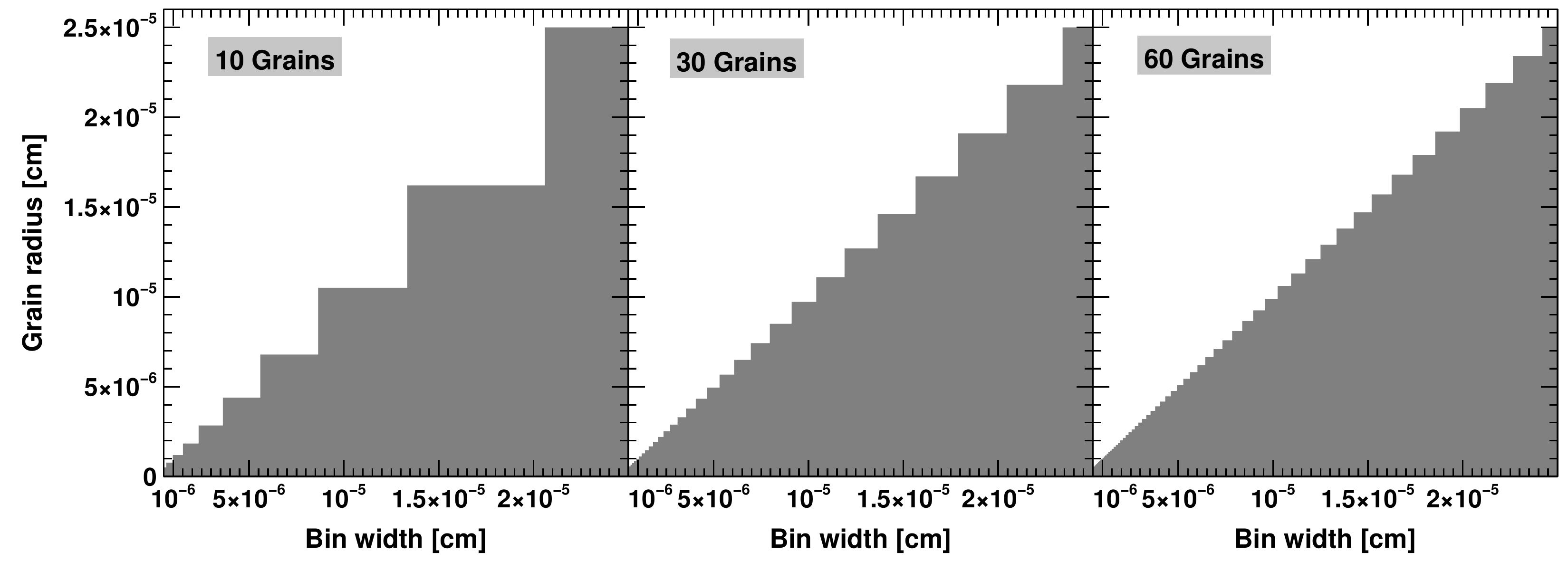}
   \caption{Bin width as a function of grain radius. The width of each vertical column represents bin width and the height of each vertical column represents the radius of the representative grain in that bin. Since we divide the entire range on log scale with equal spacing, in linear scale the width of each bin is different and increases with radius except for the last bin.}
   \label{Fig:grainBin}
\end{figure*}
\begin{figure}
   \includegraphics[width=.47\textwidth,trim = 0cm 0cm 0cm 0cm, clip,angle=0]{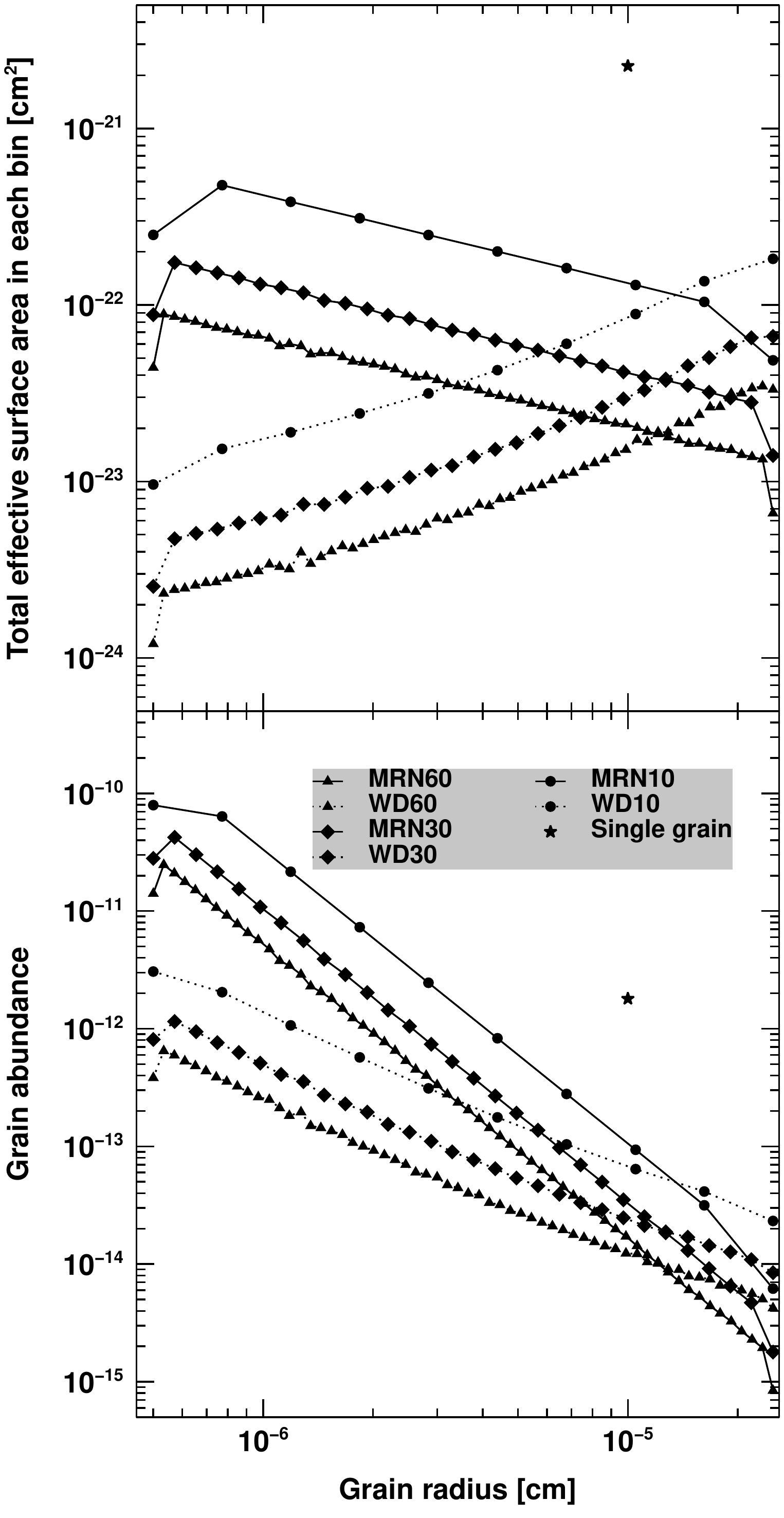}
   \caption{Bottom panel, the grain abundance (with respect to the total proton density) in each bin plotted against the grain radius. Top panel,   the total effective grain surface area (in cm$^2$)  of each bin for different models plotted against the grain radius. Legends apply to both panels.}
   \label{Fig:grainAbn}
\end{figure}
In the original version, grains are considered to have a single size of 0.1$\mu$m and a silicate composition (\cite{Hasegawa1992}). Using 0.1$\mu$m as the grain radius, assuming a gas to dust mass ratio of $10^2$ and a dust bulk density of 3 g.cm$^{-3}$, we obtain a grain number density of $1.8\times 10^{-12}\ n_{\rm H}$, where $n_{\rm H}$ is the number density of H (in cm$^{-3}$). If there are $10^6$ binding sites on the surface of a 0.1$\mu$m grain (see \cite{Hasegawa1992}), we then obtain a surface site density of $8 \times 10^{14}$~cm$^{-2}$. These values are the ones usually used in chemical models and also in our "single grain size" model.

The aim of this work is to include several sizes of grains in Nautilus to obtain a more realistic description of the ISM conditions. This modified version of Nautilus is called Nautilus Multi Grain Code (NMGC). To obtain the different grain sizes, we have sampled the total range between the smallest and the biggest grains ($0.005\mu$m and $0.25\mu$m, respectively) in our simulation in either 10, 30, or 60 bins. These numbers are arbitrary and one can use any desired number of grain sizes in the simulations provided computational time permits. For example, on a laptop with Intel\textsuperscript{\textregistered} Xeon\textsuperscript{\textregistered} E3-1505M v5 CPU @ 2.80GHz, a simulation with ten grain sizes or bins takes about 2 mins to reach $10^7$ yr but with 60 grain sizes, to reach this point takes about 30 mins. 

In NMGC, all grains are connected with each other through a common gas phase. Two grains never interact directly or exchange mass or energy directly with each other. In other words there is no collision of dust grains. Exchange of mass is only possible through the process of desorption of the surface species to the gas phase and the accretion of gaseous species on the dust surface.   However, one side effect of the rate equations method is that all species on the grains of the same size can interact with each other (as was the case for the single grain size model also) but there is no
interaction between surface molecules on grains of different sizes. In our multi-grain model, the chemical networks on the surface of grains have been duplicated for different bins by adding a number as prefix to each grain species. This number is specific to each bin.
Some or all major properties of each grain size in the multi-grain network can be varied independently to meet the specific requirement of various environments. The most important properties of a grain in regards to modeling a chemical evolution are the radius and corresponding grain number density, the surface temperature, and the grain composition. To compute the number densities of each grain bin, we used two grain size distribution models. The first model is the MRN distribution from \cite{Mathis1977}. The MRN model gives the number density of grains with radii between $r$ and $r + dr$ by the expression:
\begin{equation}
 \frac{1}{n_{\rm{}H}} \frac{dn_{gr}}{dr} = Cr^{-3.5};\ \ r_{min} < r < r_{max},
 \label{eqn:MRN}
\end{equation}
where $n_{gr}$ is the grain number density, $r$ is grain radius, $n_{\rm H}$ is H number density and $C$ is called grain constant. $C$ is given by $10^{-25.11}$ cm$^{2.5}$ for silicate grains and $10^{-25.13}$ cm$^{2.5}$ for graphite grains (\cite{Weingartner2001}). The above relation is valid between $r_{min} = 0.005\ \mu$m and $r_{max} = 0.25$ $\mu$m. The second model we used is widely called the WD model  from \cite{Weingartner2001}. The WD model relies on many parameters that are specific to different regions of the ISM. Following \cite{Acharyya2011}, we selected $R_v = 5.5$, case A, and $b_c = 3 \times 10^{5}$. These parameters are suitable for dense clouds. 

The total grain surface available for accretion of the species directly depends on the number density of grains. The total effective surface area depends on our choice of sampling (number of bins) as well as on the grain size distribution model. In previous works, \cite{Acharyya2011} used only five different sizes of grains in their multi-grain model, \cite{Pauly2016} used between five and eleven bins, and \cite{Ge2016} used nine different bins. \cite{iqbal2014} showed that if the number of different grain sizes is less than 30, then the total integrated grain density is significantly smaller as compared to the total grain density predicted by the grain size distribution models. In this work, we have considered 1, 10, 30, and 60 grain sizes, and studied the impact of this choice on the chemistry. Another advantage of using a better sampling is that we obtain a better estimate of each grain family temperature as a function of size.

The MRN model is valid between $r_{min} = 0.005\ \mu$m and $r_{max} = 0.25$ $\mu$m. We used the same interval for both the MRN and WD models for a better comparison of the results. We divided the range ($0.005\ \mu$m to $0.25$ $\mu$m) into 10, 30 and 60 logarithmically equally spaced points, which gives us a value for the radius $r_i$. We use this $r_i$ value to calculate the surface area and other surface parameters for the corresponding range defined by $r_{min} = (r_{(i-1)}+r_i)/2$ and $r_{max} = (r_{(i+1)}+r_i)/2$, except for the first bin in which $r_{min} = r_i$ and last bin in which $r_{max} = r_i$. Then we calculate the grain number density $n_{gr(i)}$ for both  models, with local $r_{min}(i)$ and $r_{max}(i)$ values. We plot the grain radius against bin width in Fig. \ref{Fig:grainBin}. The height of each rectangular column  represents the grain radius $r_i$ and the width of each rectangular column represents the bin width or  $r_{max}(i) - r_{min}(i)$. In Figure \ref{Fig:grainAbn}, we plot the integrated grain number density or grain abundance as a function of $r_i$ (bottom panel) and the effective surface area ($n_{gr(i)} \times \ 4\pi r_i^2$) as a function of $r_i$ (top panel). 
\begin{table}
   \begin{center}
   \caption{Total effective grain surface area   available for accretion (per cm$^3$ of space) in different models.}
   \label{tbl:T-eff_area}
   \begin{tabular}{@{}ll}
   \hline
   \hline
 Model               &   Total effective surface area [cm$^2$] \\ 
 \hline \hline
 Single grain size    &   $2.26\times 10^{-21}$ \\
\hline
 MRN 10 grains       &   $2.31\times 10^{-21}$ \\     
 MRN 30 grains       &   $2.36\times 10^{-21}$ \\     
 MRN 60 grains       &   $2.37\times 10^{-21}$ \\ 
 \hline
 WD  10 grains       &   $6.10\times 10^{-22}$ \\   
 WD  30 grains       &   $6.32\times 10^{-22}$ \\  
 WD  60 grains       &   $6.37\times 10^{-22}$ \\  
\hline
   \end{tabular}
   \end{center}  
\end{table}

In Table \ref{tbl:T-eff_area}, we show the total effective surface areas in cm$^2$ (available for accretion per cm$^3$ of space) as calculated using different grain size distribution models. It is to be remembered that the total effective surface area depends on the bin size as well as on the radius used to represent that bin. We have tried to ensure that the total effective surface areas are close to each other in the same grain size distribution cases but with a different number of bins. This is to minimize the possible effects on the accretion rates of different species caused by using   different numbers of bins. It can also be seen that some how the total effective surface areas calculated using the MRN models are very close to what we get in the single grain model in which the number density of grains is calculated by the dust to gas mass ratio only. The total effective surface area in the WD model is almost three and half time less than that in other models. This is because, in comparison to the MRN model, in the WD model   small  grains are less abundant and   big  grains are more abundant (see Fig. \ref{Fig:grainAbn}, bottom panel) and surface area decreases as we add   small  grains to make   big  grains, keeping the total mass constant.

\subsection{Grain surface temperature and cosmic-ray-induced desorption}
\label{M_sec:1}
\begin{figure}
    \centering
   \includegraphics[width=.47\textwidth,trim = 0cm 0cm 0cm 0cm, clip,angle=0]{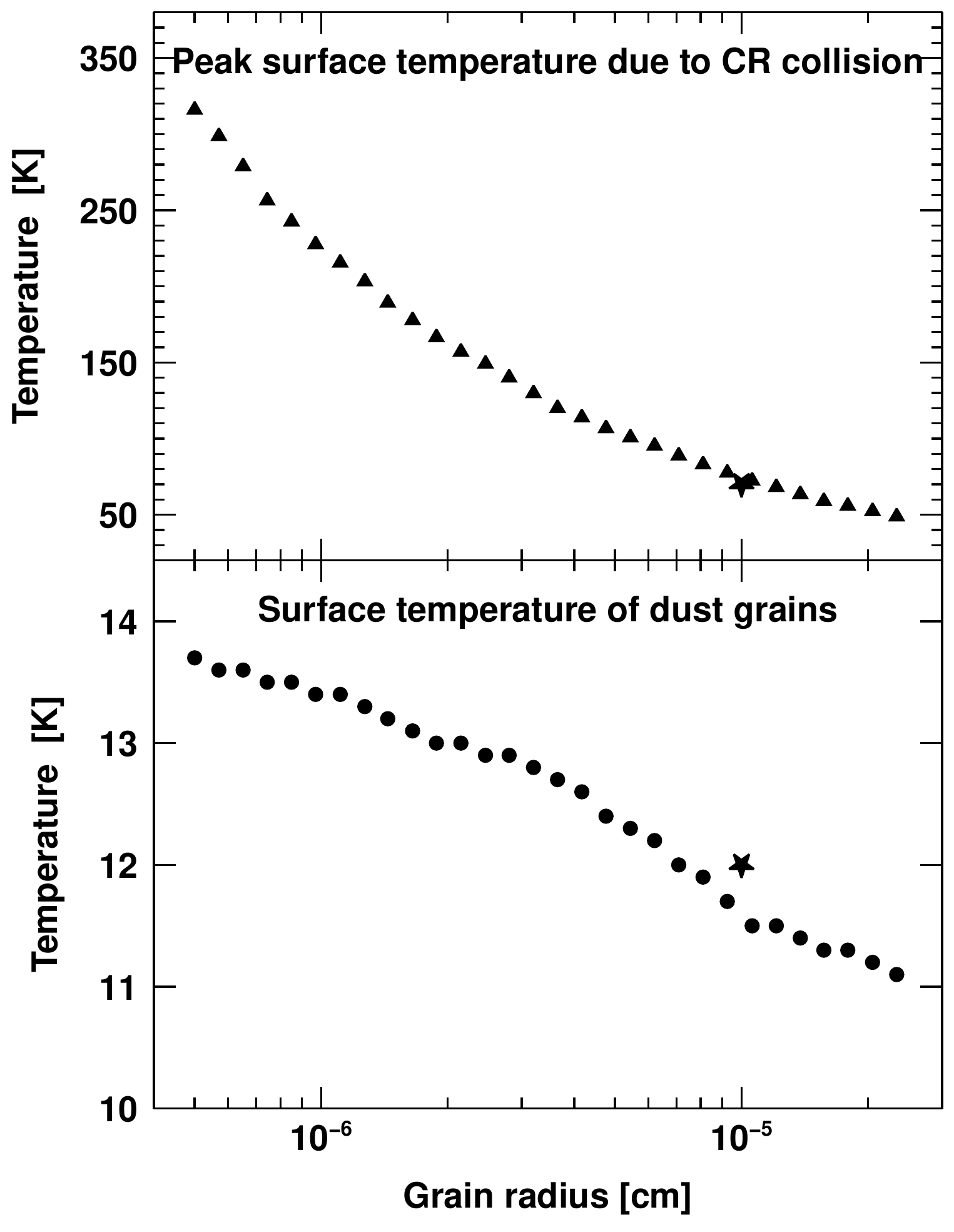}
   \caption{Bottom panel: grain temperature (circles for multi-grain model and star for single-grain model) as a function of grain radius. Top panel: Peak surface temperature (triangles for multi-grain model and star for single-grain model) due to collision with cosmic rays as a function of grain radius.} 
   \label{Fig:grainTemp}
\end{figure}
For the surface temperatures of different grain sizes, we assumed two different sets of models. First, we kept the grain temperature constant at 10K for all grain sizes. As a second case, we used a different temperature for each bin or grain size. To derive this temperature, we first compute the steady state dust temperature in diffuse medium conditions as a function of grain radius from \cite{Draine1984}. Then we followed the recommendation from \cite{Mathis1983} to multiply this value by $0.75$ for dense cold medium. For our single-grain-size model, this method gives a dust temperature of 12K. In our model we did not consider fluctuation in dust temperature due to UV or cosmic rays, but cosmic-ray-induced desorption is implemented in Nautilus code following the method described in \cite{Hasegawa1993}. We modified our multi-grain model to account for different grain sizes. The cosmic-ray-induced desorption rate for any species on the $i$th grain is given by the equation:
\begin{equation}
 k_{crd}(i) = f(T_{max}(i))k_{evap}(i,T_{max}(i))
 \label{crdr}
,\end{equation}
where $f(T_{max}(i))$ is the duty cycle of the $i$th grain at elevated temperature  $T_{max}(i)$ and $k_{evap}(i,T_{max}(i))$ is the evaporation or thermal desorption rate for the species on the $i$th grain at temperature $T_{max}(i)$. To estimate $T_{max}(i)$, we used the cosmic-ray peak temperature as calculated by \cite{Herbst2006} as a function of grain radius (assuming we have only silicate grains). $f(T_{max}(i))$ is defined as the ratio of time-scale for cooling via desorption of volatiles to the average time interval between two successive cosmic ray hits. Following \cite{Hasegawa1993}, for 0.1~$\mu$m grains, this interval is taken to be equal to $3.16 \times 10^{13}$ s,  which comes from the Fe cosmic ray flux in \cite{Leger1985}. We simply scaled it to obtain the flux for different grain sizes.   At the densities of cold cores, cosmic rays are not completely attenuated \citep{Umebayashi1981}. The cosmic rays deposit only a part of their energy into a grain during the interaction (see Eq. (4) of  \cite{Herbst2006}).

In previous models with single classical grains, the time-scale for cooling via desorption is taken to be $10^{-5}$ s. This time is the half life of CO on the dust surface at 70K. It is assumed that when a grain is heated to 70K by collision with a cosmic-ray particle, CO is the most prominent species to desorb from the surface causing the grain cooling. It is also assumed that within this fraction of time, about $10^6$ species would desorb from the surface, taking with them the extra energy deposited by cosmic rays and resulting in significant cooling of the grain.  But when we are considering   small  grains of radii as small as 0.005~$\mu$m with a total number of sites equal to 2500 only, it is obvious that on   small  grains most of the time surface population would be only a fraction of $10^6$. The cooling rates of   small  grains are then expected to be much slower and should result in lower surface population. In addition, this formalism of the cosmic ray desorption assumes that there are CO molecules on the surface. In the absence of a better method to calculate or to scale this time for different grain sizes we used the same time period of  $10^{-5}$ s for all grain sizes in our multi-grain model. Considering the importance of this assumption, further investigation in this process is needed.

The bottom panel of Fig. \ref{Fig:grainTemp} shows the grain radius for different models and corresponding dust temperature, while the top panel shows the cosmic ray peak temperature as a function of grain radius. 
\subsection{Other model parameters}
\begin{table}
   \begin{center}
   \caption{Elemental abundances and initial abundances.}
   \label{tbl:initial-abundances}
   \begin{tabular}{@{}lll}
   \hline
   \hline
   Element                      &       Abundance relative to H    & References  \\
   \hline
   H$_2$                        &       0.5                        &             \\
   He                           &       0.09                       &$^\textrm{a}$ \\
   N                            &       6.2$\times10^{-5}$         &$^\textrm{b}$ \\
   O                            &       2.4$\times10^{-4}$         &$^\textrm{c}$ \\      
   C$^+$                        &       1.7$\times10^{-4}$         &$^\textrm{b}$ \\   
   S$^+$                        &       8.0$\times10^{-9}$         &$^\textrm{d}$ \\      
   Si$^+$                       &       8.0$\times10^{-9}$         &$^\textrm{d}$ \\
   Fe$^+$                       &       3.0$\times10^{-9}$         &$^\textrm{d}$ \\
   Na$^+$                       &       2.0$\times10^{-9}$         &$^\textrm{d}$ \\
   Mg$^+$                       &       7.0$\times10^{-9}$         &$^\textrm{d}$ \\   
   P$^+$                        &       2.0$\times10^{-10}$        &$^\textrm{d}$ \\      
   Cl$^+$                       &       1.0$\times10^{-9}$         &$^\textrm{d}$ \\ 
     ice                     &       0                          &              \\
   \hline
   \end{tabular}
   \end{center} 
   $^\textrm{a}$\cite{Wakelam08}, $^\textrm{b}$\cite{Jenkins09}, 
   $^\textrm{c}$\cite{Hincelin2011}, $^\textrm{d}$\cite{Graedel82}   
\end{table}
\begin{table}
   \begin{center}
   \caption{Some important parameters used in our models.}
   \label{tbl:cold_dense_cloud_model}
   \begin{tabular}{@{}lll}
   \hline
   \hline
   Parameters                     & Value\\
   \hline
   $T_{gas}$                      & 10 K \\
   $n_\textrm{H}$                 & $2\times10^4$ cm$^{-3}$ \\
   $A_V$                          & 15 \\
   Cosmic rays ionization rate    & $1.3\times10^{-17}$ s$^{-1}$ \\
   Grain surface site density     & $8.0\times10^{14}$  cm$^{-2}$   \\
   Initial abundances             & see Table \ref{tbl:initial-abundances} \\
   \hline
   \end{tabular}
   \end{center}  
\end{table}
We have run the different models using parameters for cold core conditions. In Table \ref{tbl:initial-abundances}, we have listed the elemental abundances and initial abundances used in our all models while Table~\ref{tbl:cold_dense_cloud_model} summarizes some important parameters which are kept the same in all models.   We start simulations with zero surface abundance.
\section{Results}
\label{results}
We divide our simulations into four sets according to the number of grain sizes used in the model. These sets have 1, 10, 30 and 60 different grain sizes. Subsequently we divide each set into various cases to gain insight into the effect of the different parameters. These cases are based on 1) including or not cosmic-ray-induced desorption, 2) the MRN or the WD grain size distribution models, and 3) the uniform or the grain-size-dependent dust surface temperature. The results of all these cases are shown and compared in this section. 
\subsection{Grain size and cosmic-ray-induced desorption}
\label{R_sec:1}
\begin{figure}
    \centering
   \includegraphics[width=.47\textwidth,trim = 0cm 0cm 0cm 0cm, clip,angle=0]{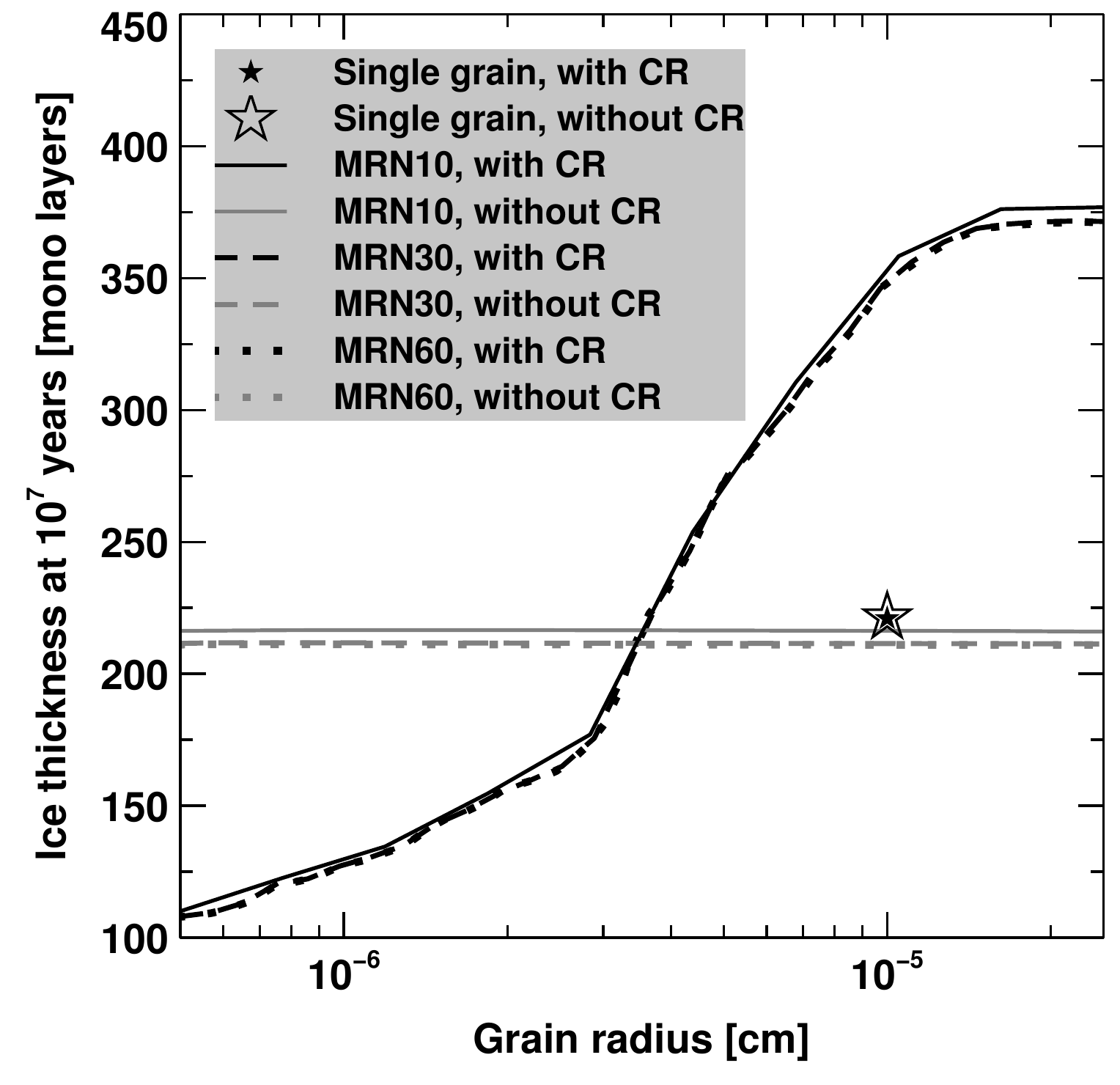}
   \caption{Ice thickness as a function of grain radius for models with the MRN distribution and with the cosmic-ray-induced desorption enabled (black lines) and disabled (gray lines). Stars show the results for the single grain model. The dust temperature is kept constant at 10K for all grains.}
   \label{Fig:2b}
\end{figure}
\begin{figure*}
    \centering
   \includegraphics[width=.97\textwidth,trim = 0cm 0cm 0cm 0cm, clip,angle=0]{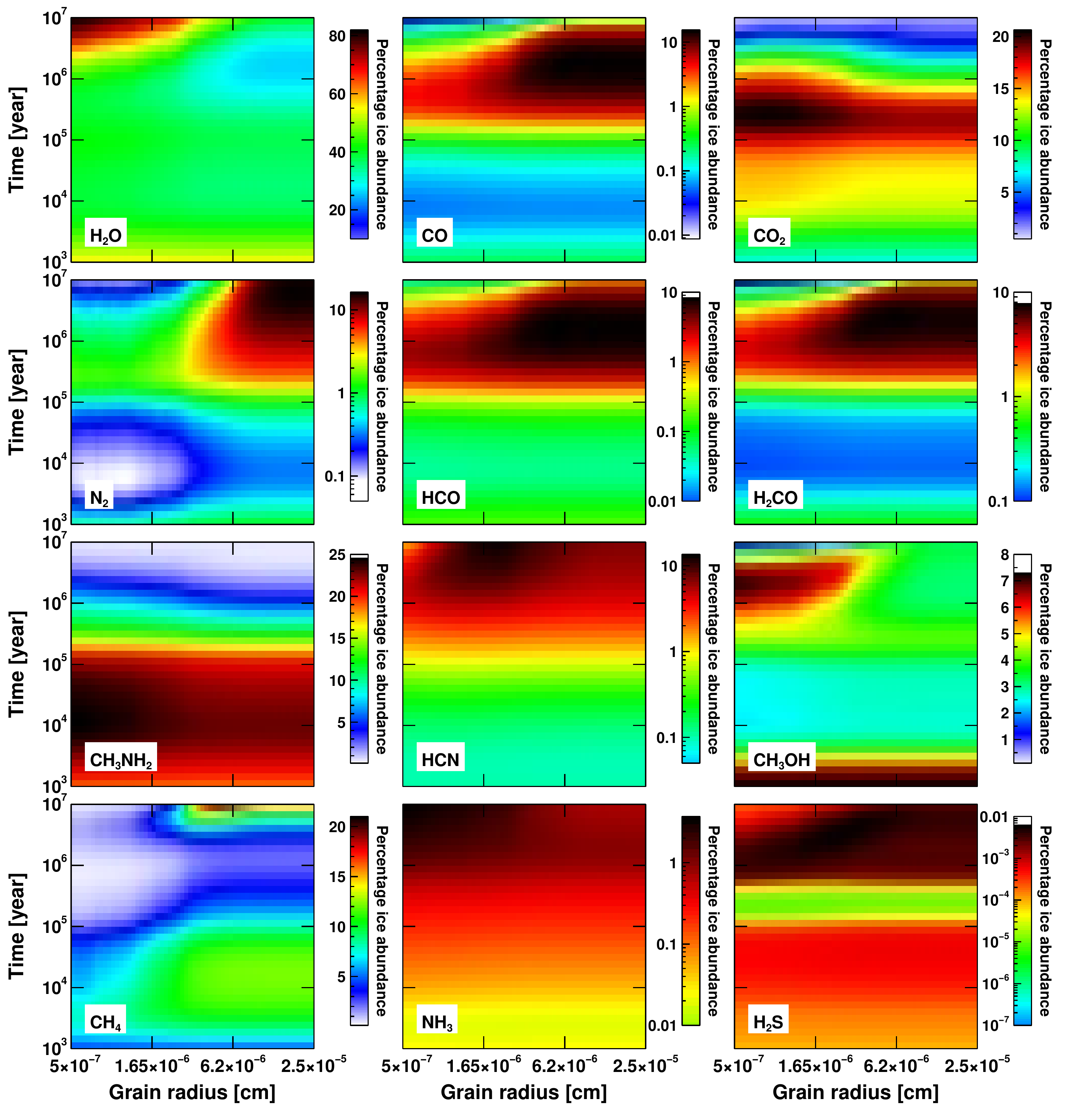}
   \caption{Percentage ice abundance of the most abundant species as a function of grain radius  (x-axis) and time (y-axis). Results are for simulation with cosmic-ray-induced desorption enabled and for the MRN grain size distribution with 60 grain sizes. The dust temperature is kept constant at 10K for all grains. }
   \label{Fig:Per_ab_map}
\end{figure*}
\begin{figure*}
   \centering
   \includegraphics[width=.97\textwidth,trim = 0cm 0cm 0cm 0cm, clip,angle=0]{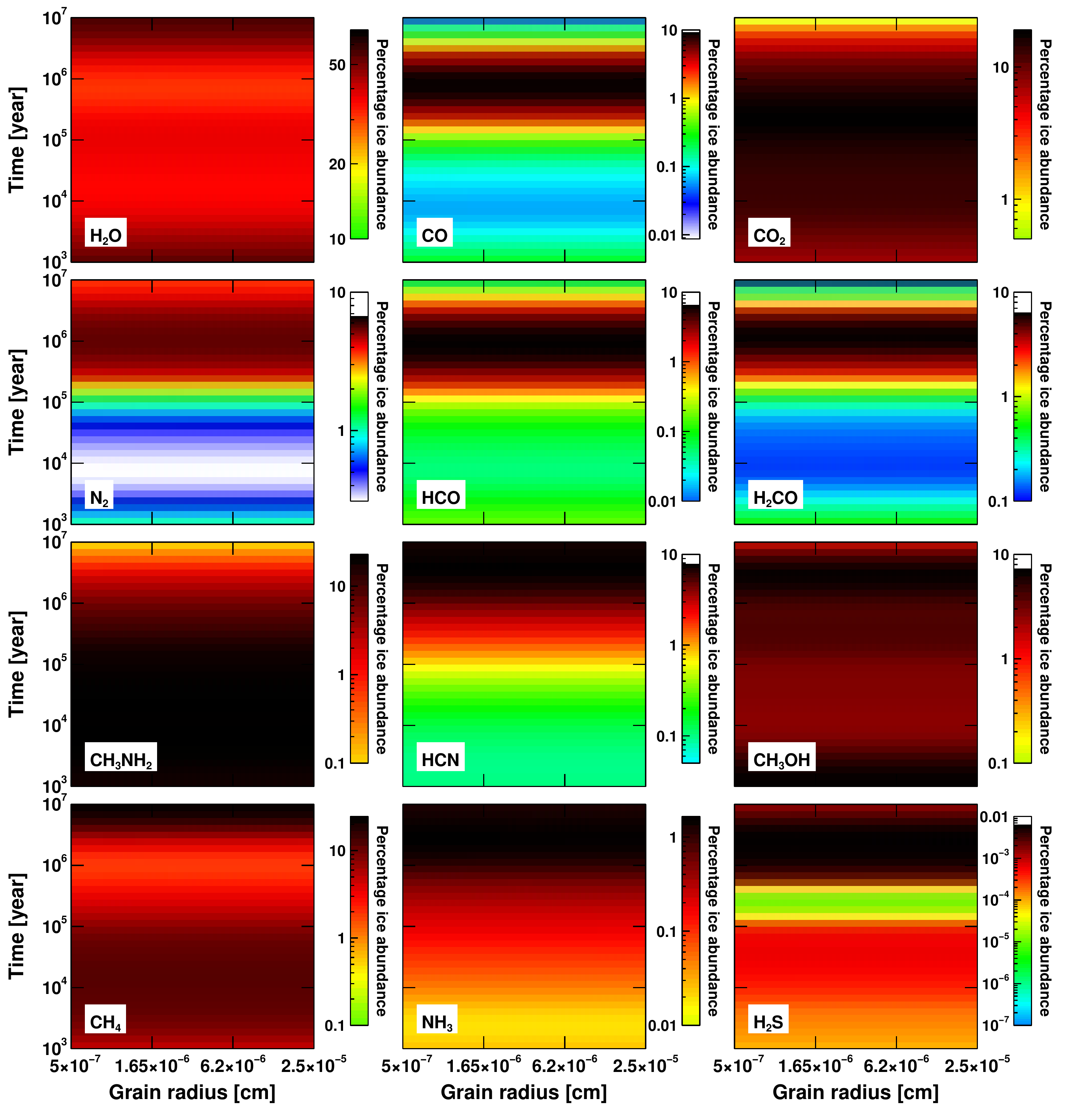}
   \caption{As in Fig.~\ref{Fig:Per_ab_map} but without cosmic-ray-induced desorption.}
   \label{Fig:Per_ab_map_noCR}
\end{figure*}
\begin{figure*}
    \centering
   \includegraphics[width=.97\textwidth,trim = 0cm 0cm 0cm 0cm, clip,angle=0]{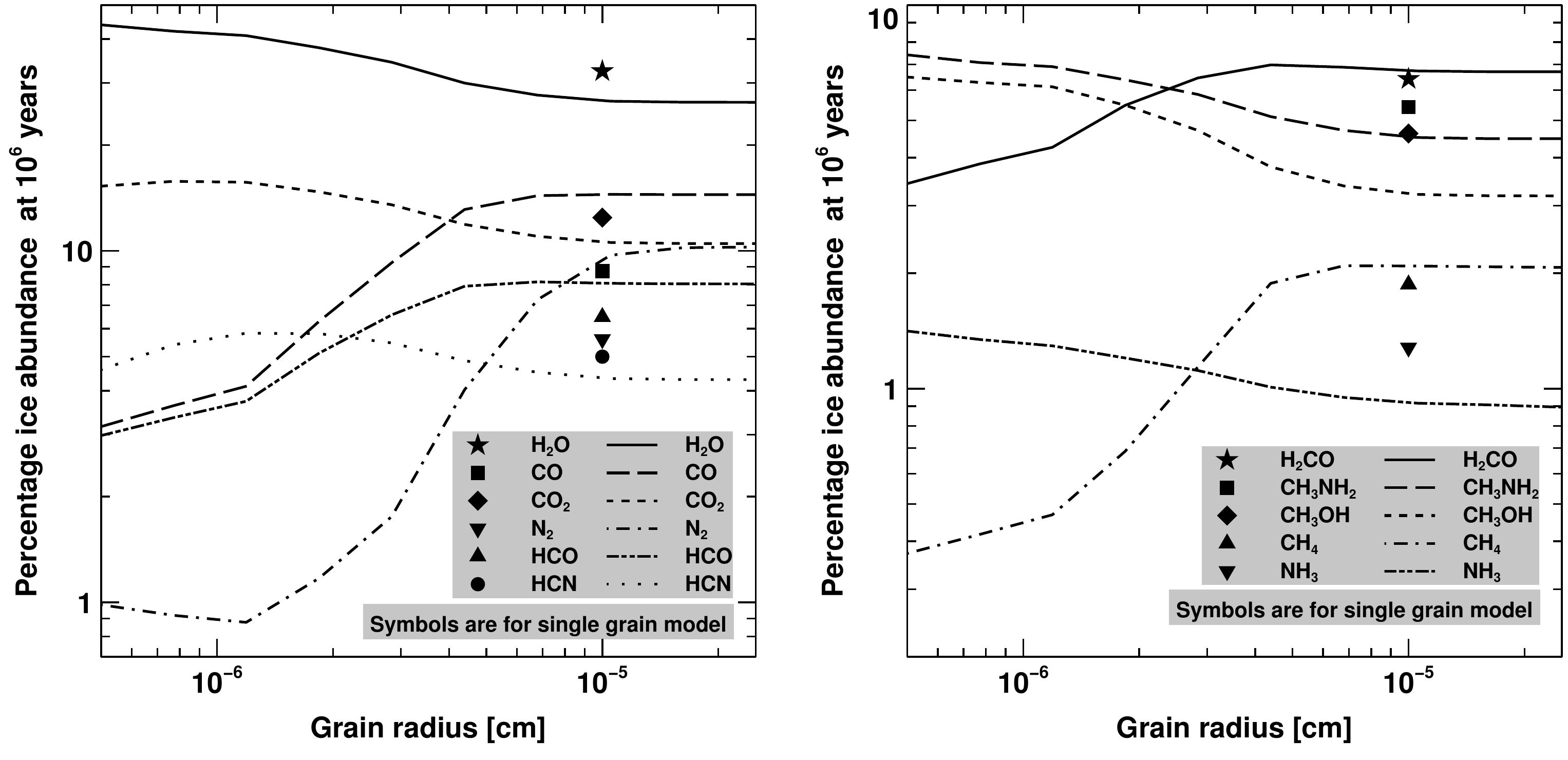}
   \caption{Percentage ice abundances of the most abundant species as a function of grain radius at $10^6$ yr. Results are for the simulation with the cosmic-ray-induced desorption enabled and for the multi-grain model (lines), with 60 grain sizes and the MRN grain size distribution, and the single-grain model (symbols). The dust temperature is kept constant at 10K for all grains.}
   \label{Fig:Per_ab_all_grain}
\end{figure*}
\begin{figure*}
    \centering
   \includegraphics[width=.97\textwidth,trim = 0cm 0cm 0cm 0cm, clip,angle=0]{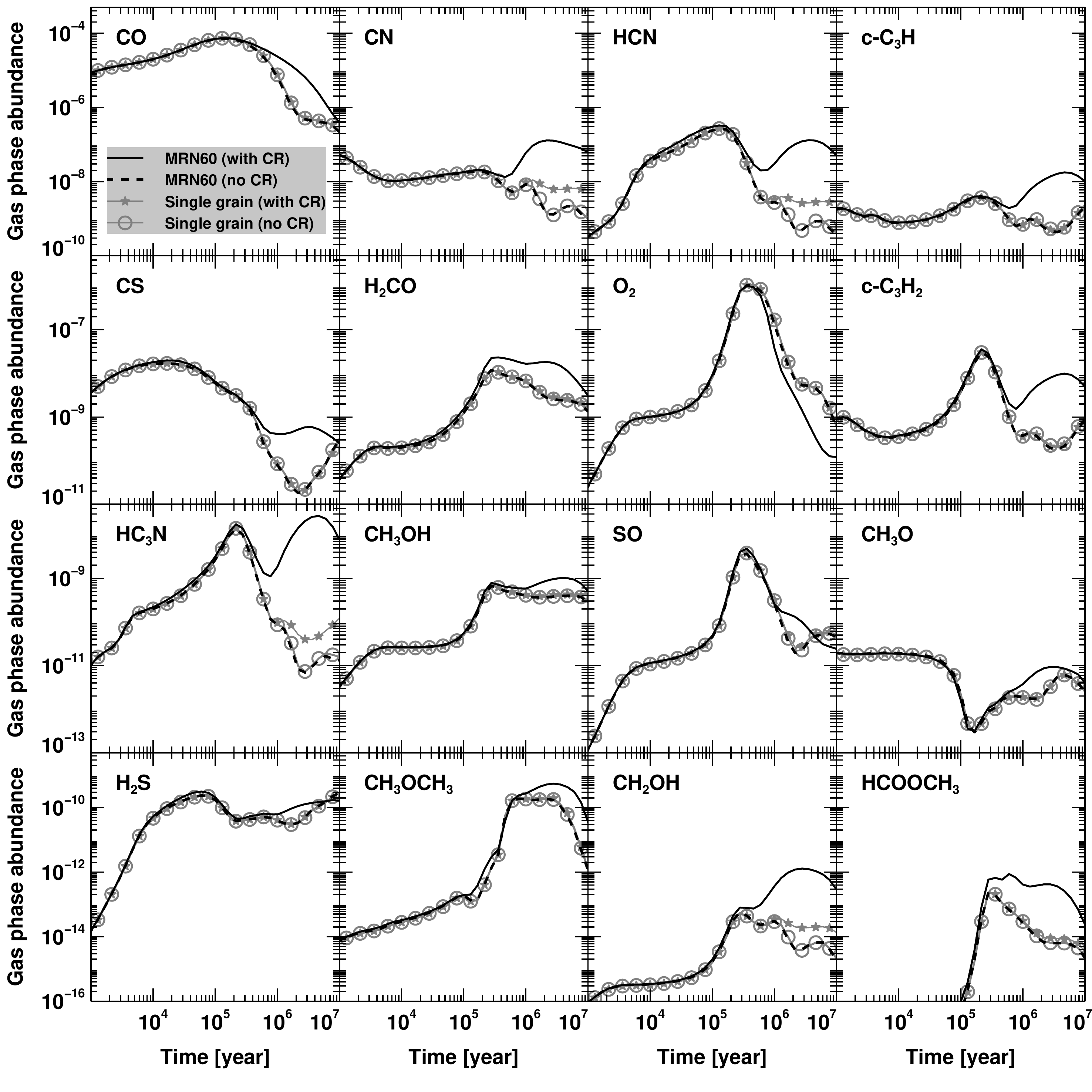}
   \caption{Gas phase abundances of selected species as a function of time. Solid and dashed black lines are for the multi-grain models with the cosmic-ray-induced desorption enabled and disabled, respectively. Gray lines with stars and circles, respectively, show the results from the single-grain model with and without the cosmic- ray-induced desorption. The dust temperature in each model is kept constant at 10K for all grains. Legends apply to all panels.}
   \label{Fig:9}
\end{figure*}
\begin{figure}
    \centering
   \includegraphics[width=.47\textwidth,trim = 0cm 0cm 0cm 0cm, clip,angle=0]{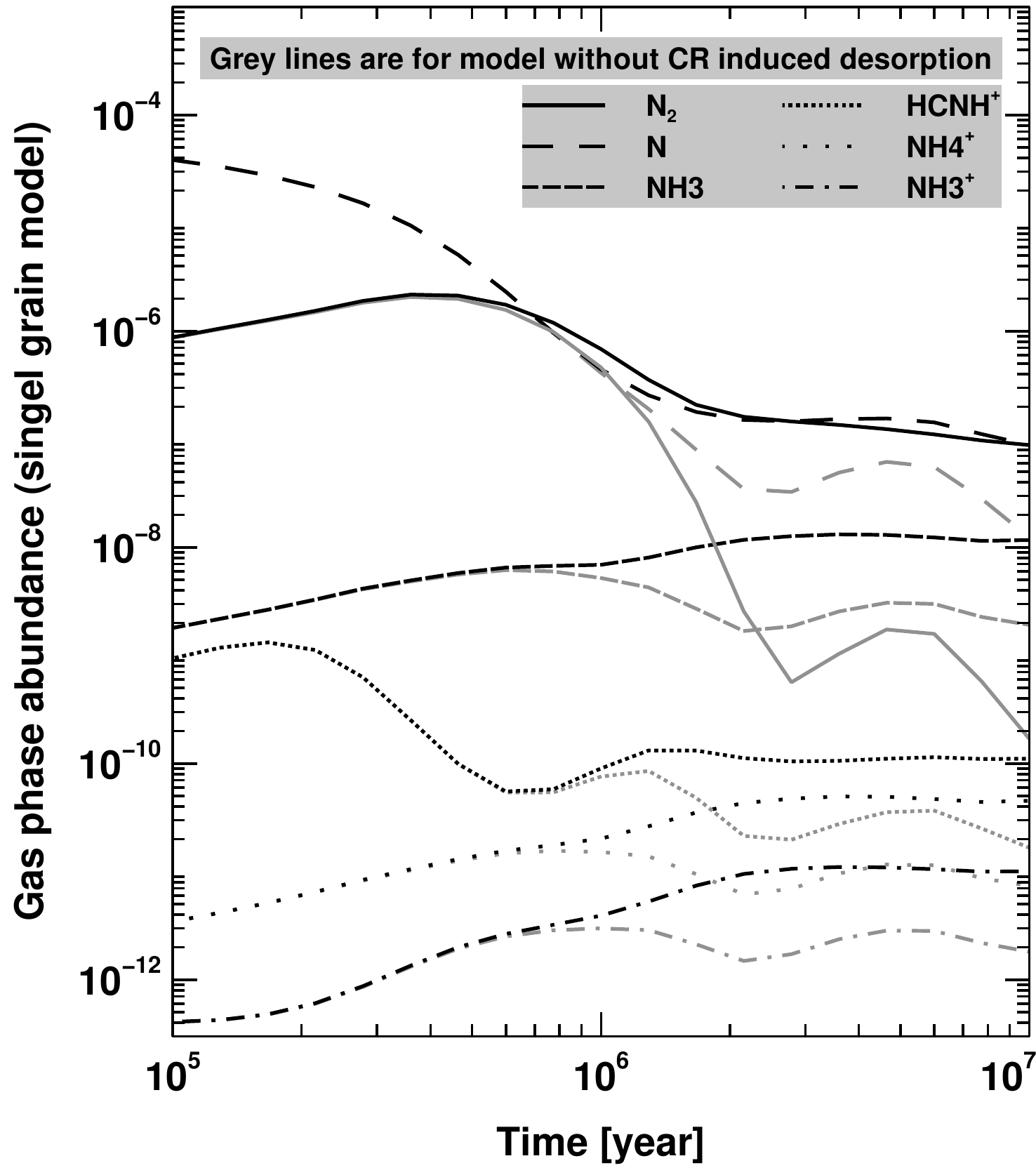}
   \caption{Simulated results of gas phase abundances of selected N-bearing species as a function of time as obtained in the single-grain model. Black and  gray lines are for models with  and without cosmic-ray-induced desorption, respectively. Dust temperature is kept constant at 10K in both models.}
   \label{Fig:N_species}
\end{figure}
\begin{figure}
    \centering
   \includegraphics[width=.47\textwidth,trim = 0cm 0cm 0cm 0cm, clip,angle=0]{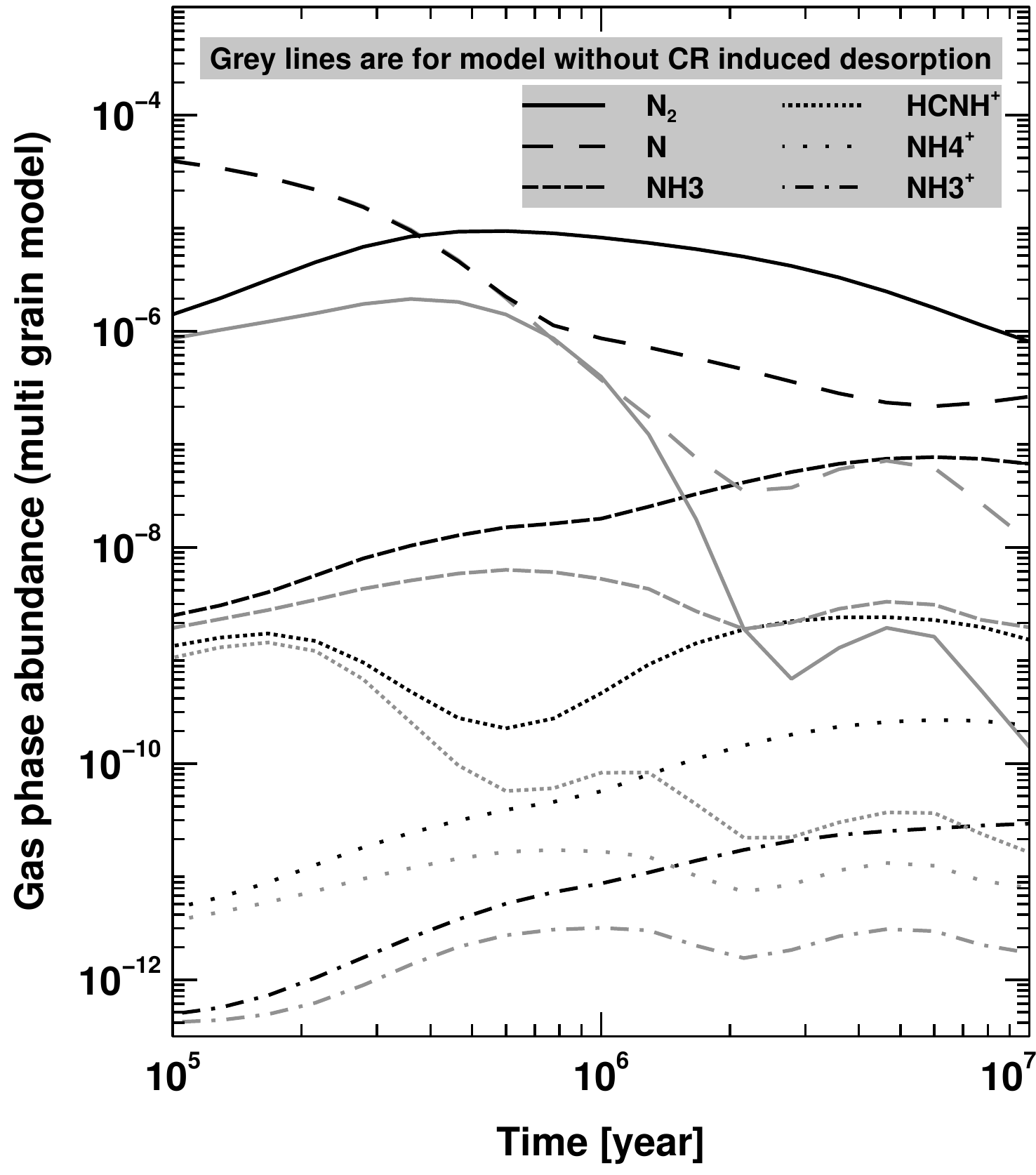}
   \caption{Gas phase abundances of selected N-bearing species as a function of time as obtained in the multi-grain model with 60 bins and with the MRN grain size distribution. Black and  grey lines are for the models with and without the cosmic-ray-induced desorption, respectively. The dust temperature is kept constant at 10K in both models.}
   \label{Fig:N_species_MGC}
\end{figure}
First, we show the importance of the cosmic-ray-induced desorption, a process in which the desorption efficiency strongly depends on the size of the grains. In Fig. \ref{Fig:2b}, we plot the computed ice thickness at $10^7$ yr as a function of grain radius for models with 10, 30, and 60 grain sizes distributed using the MRN grain size distribution model. We also show the same for the single-grain model. The temperature of the dust grains is kept constant at 10K in all models. When we do not include the cosmic-ray-induced desorption process, the ice thickness is almost the same on all the grain sizes, and this thickness is very close to what we get in the single grain model. When the cosmic-ray-induced desorption is included, we see that the   small  grains have lost mass while the   big  grains have gained mass as compared to the previous case.
The loss of mass from the   small  grains can be explained by the fact that the rise in the surface temperature, due to a cosmic ray particle hitting the dust grain, can cause the desorption of about $10^6$ lighter molecules, such as CO (see \cite{Hasegawa1993}), before the dust is significantly cooled.  Since the   small  grains have a very small number of binding sites (e.g., the smallest grain in our simulation has only 2500 binding sites), desorption of $10^6$ molecules from the surface may result in almost complete destruction of the surface ice unless there are a few hundred mono-layers of ice or if the ice is mostly composed of more strongly bounded species.
The   big  grains gain mass by accretion of these species back to the grain surface. Although   big  grains also lose mass due to the cosmic-ray collisions, this   loss  is a very small portion of the total surface mass (a   loss  of $10^6$ CO molecules is a   loss  of less than one mono layer of ice for   big  grains) as the surface area of the   big  grains is huge. This is also the reason why we notice almost no change in the surface population in the single grain model due to the cosmic-ray-induced desorption process although the lighter molecules such as CO and N$_2$ are constantly evaporated due to collisions with cosmic rays. This is also because the evaporated species in the single grain model come back to recondense the same grain. 

In Fig. \ref{Fig:2b}, we can see that the mass loss on different   grains  is not linear with the grain size. Furthermore, we see three regimes. From $0.25\mu$m to $0.1\mu$m, the mass gain is rather constant, there is a relatively fast loss of mass between the grains of $0.1\mu$m to $0.03\mu$m in radius, and this rate then slows down again for grains smaller than $0.03\mu$m.  These variations are very likely due to some threshold effects on the desorption temperatures of abundant molecules. For the big grains (radius larger than $0.1\mu$m), the cosmic-ray-induced peak temperature is smaller than 70K. At these temperatures only light species such CO can desorb. Between $0.03\mu$m and $0.1\mu$m, the peak temperature is between 70 and 140~K. At these temperatures, most of the ice constituents can desorb. As a consequence, on grains smaller than $0.03\mu$m, only species with the strongest bounds can remain. These species would require a much higher peak temperature to desorb.  

This mass transfer may result in a significant change in the ice composition depending on grain size. In Fig. \ref{Fig:Per_ab_map}, we plot colored maps showing the percentage of ice abundance for selected species, with respect to the total ice abundance.
For comparison, we show in Fig. \ref{Fig:Per_ab_map_noCR} the same maps but without the cosmic-ray-induced desorption. To complete the analysis, we show a simple line plot (see Fig. \ref{Fig:Per_ab_all_grain}) of the percentage ice abundances of the most abundant species as a function of grain radius at $10^6$~yr. In this figure, we also show the percentage ice abundance obtained with the single-grain-size model. The chemical composition on different grain sizes is indeed different when cosmic-ray-induced desorption is included, while it remains constant with grain size without it. 

In general, the selected species can be divided into two groups by looking at their trends. In the first group we have CO, N$_2$, HCO, H$_2$CO and CH$_4$, and in the second group we can put all other species plotted here. The basic trend of species in the first group is that they have a lower percentage abundance on the   small  grains compared to the   big  grains. We also note that desorption energies of all these species are between 1100K to 2100K. In the second group we see a completely opposite trend, that is, the percentage abundance of all species is larger on the   small  grains and it decreases as the grain size increases. We also note that the desorption energy of all these species are above 5000K except CO$_2$ whose desorption energy is 2575K. Another interesting point is that the percentage abundance as obtained in the single-grain model (see symbols in Fig. \ref{Fig:Per_ab_all_grain}) for the species of the first group is lower than that in the multi-grain model for the same size grain (0.1$\mu$m) while it is opposite for all species in the second group. 
 
In Fig. \ref{Fig:9}, we show the gas-phase abundances of selected species for the above simulations. We see that the results with the single grain model, with or without the cosmic-ray-induced desorption (in Fig. \ref{Fig:9}, lines with stars and circles, respectively), are almost overlapping with the exception of N-bearing species and CH$_2$OH; for these species curves very much overlap with each other until close to $5\times10^5$ yr where they begin to diverge. All of these species have a larger gas-phase abundance in the model with cosmic-ray-induced desorption. It is to be noted that gas phase CH$_2$OH is mainly produced via CH$_3$OH + CN $\rightarrow$ CH$_2$OH + HCN reaction, and the increase in CH$_2$OH production at later times is due to an increase in abundance of CN in the gas phase. Other N-bearing species, including CN, are increased in the gas phase because of higher gas phase abundances of N and N$_2$. Our analysis showed that, in the simulation with cosmic-ray-induced desorption, at about $5\times10^5$ yr, almost 8\% of N$_2$ production is via the desorption of N$_2$ ice, while this is almost nil when the cosmic ray induced desorption is not included. This over saturation of N$_2$ in the gas phase leads to an increase in  abundances of N and N$^+$ (N$_2$ + He$^+$ $\rightarrow$ He + N + N$^+$). This finally results in an increase in abundances of various N-bearing species such as CN, HCN, and many more (see Fig. \ref{Fig:N_species}). Although we observed a noticeable impact of the cosmic-ray-induced desorption in the single-grain model for N-bearing species, for most other species the effect is not noticeable even up to 10$^7$ yr.

In the multi-grain model, the gas phase abundance in most species is increased (see solid black lines in Fig. \ref{Fig:9}) due to the cosmic-ray-induced desorption and the differences in results for the models with and without the cosmic-ray-induced desorption are even wider for N-bearing species (see Fig. \ref{Fig:N_species_MGC}). In the multi-grain model, the effect of the cosmic-ray-induced desorption is noticeable even after $10^4$ yr while in the single-grain model it is visible only after $5\times10^5$ yr. An increase in the abundance of one species can result in the decrease of another species, as can be seen with O$_2$; its gas phase abundance is reduced in the multi-grain model with the cosmic-ray-induced desorption while it is not noticeable in the single-grain model. In fact, the gas phase O$_2$ is mainly reduced due to its increased destruction via two reactions, CN + O$_2$ $\rightarrow$ O + OCN and C + O$_2$ $\rightarrow$ O + CO, as the gas phase abundances of both CN and C are increased. 

A greater abundance of species,
such as CH$_3$OH, which mainly form on the grain surface, clearly indicates the effect of a more efficient desorption from the small grains at elevated temperature, although this elevated temperature stays for a very short time (only $10^{-5}$ s) after each cosmic-ray collision before the surface temperature returns to normal. It should also be noted that the   small  grains are much higher in number (see Fig. \ref{Fig:grainAbn}), meaning that any process which affects the   small  grains will show more prominent results overall.  

To summarize, in this section, we explored the effect of the cosmic-ray-induced desorption in the single-grain and multi-grain models. We explained the differences in the ice compositions on grains of different sizes due to the cosmic-ray-induced desorption. We showed that a more efficient desorption of physisorbed species from the   small  grains results in an increase in concentrations of certain species in the gas phase which in turn increases accretion rates causing   big  grains to gain more mass. Without the cosmic ray induced desorption, we did not see any change in the ice compositions on different grain sizes \citep[in agreement with][]{Acharyya2011,Pauly2016}. We also observe that in the single grain model, the cosmic ray desorption process is a minor one for most of the species except N-bearing species so the results with or without it are very similar at least on the dust surface.

\subsection{Grain number and effect of grain size distribution}
\label{R_sec:2}
\begin{figure}
    \centering
   \includegraphics[width=0.47\textwidth,trim = 0cm 0cm 0cm 0cm, clip,angle=0]{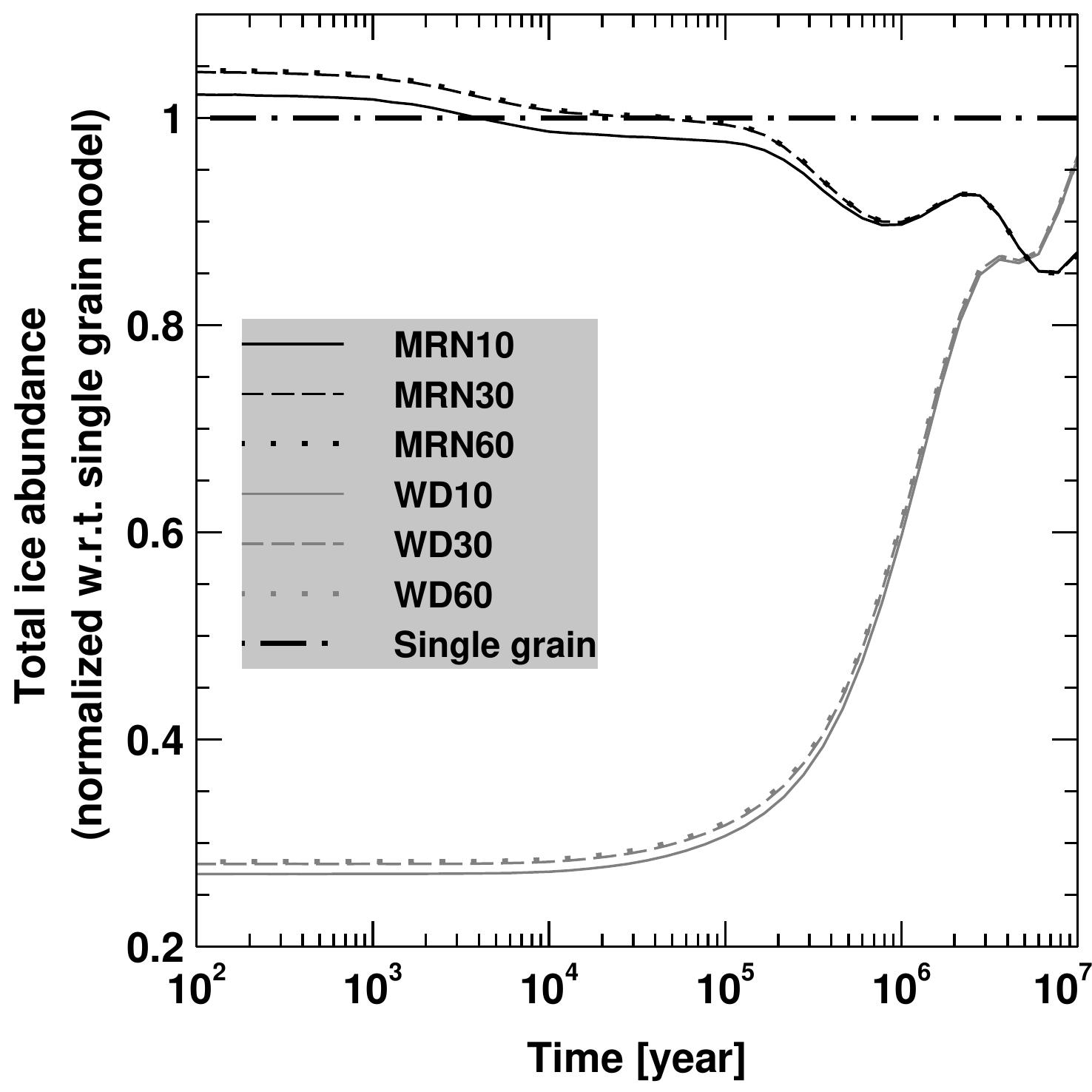}
   \caption{Normalized total ice abundances as a function of time (see text). Cosmic-ray-induced desorption was enabled and the surface temperature was kept constant at 10K in all models.}
   \label{Fig:3}
\end{figure}
\begin{figure}
    \centering
   \includegraphics[width=.47\textwidth,trim = 0cm 0cm 0cm 0cm, clip,angle=0]{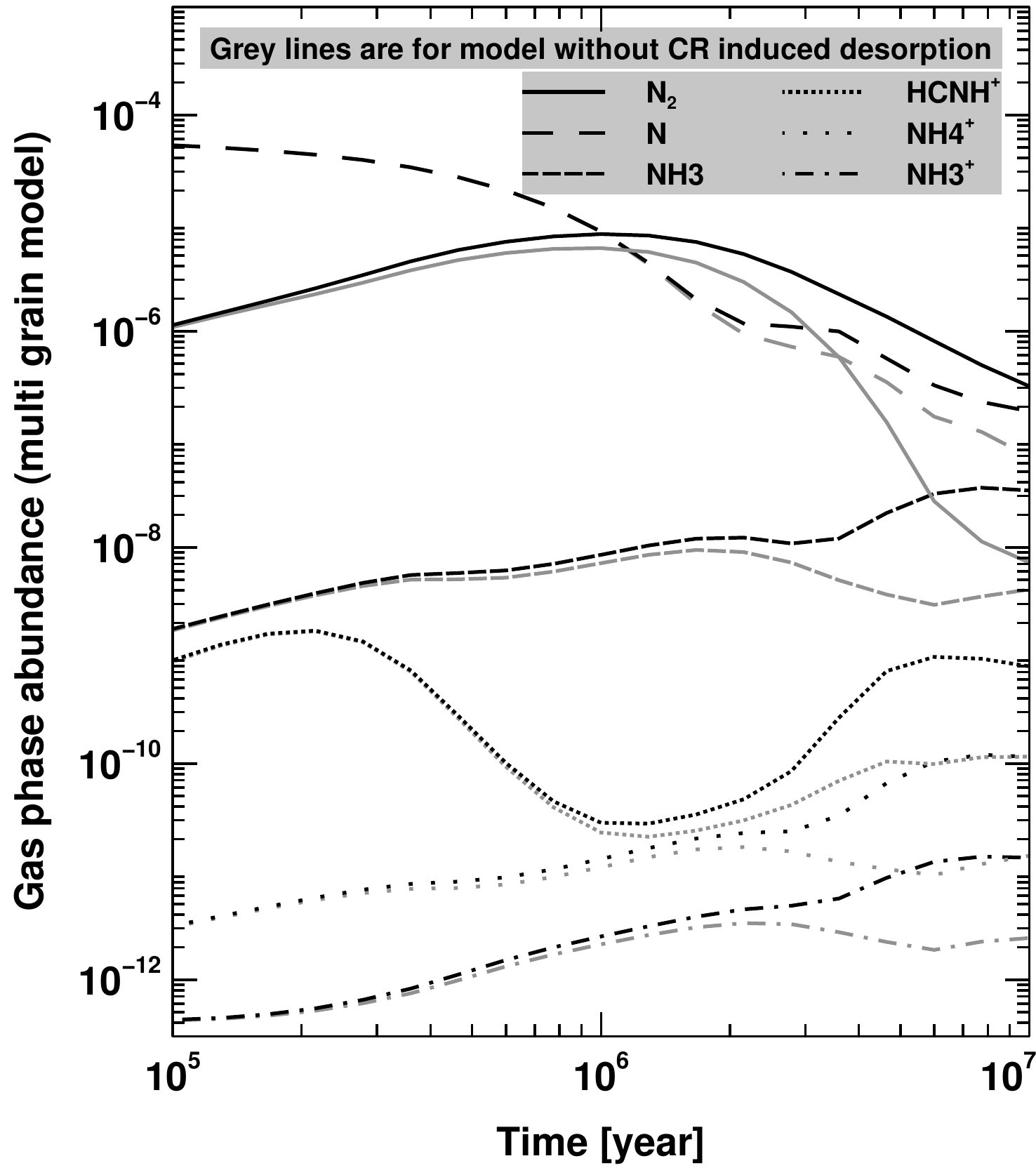}
   \caption{As in Fig. \ref{Fig:N_species_MGC} but for the simulation using the WD distribution.}
   \label{Fig:N_species_WD}
\end{figure}
\begin{figure*}
    \centering
   \includegraphics[width=.97\textwidth,trim = 0cm 0cm 0cm 0cm, clip,angle=0]{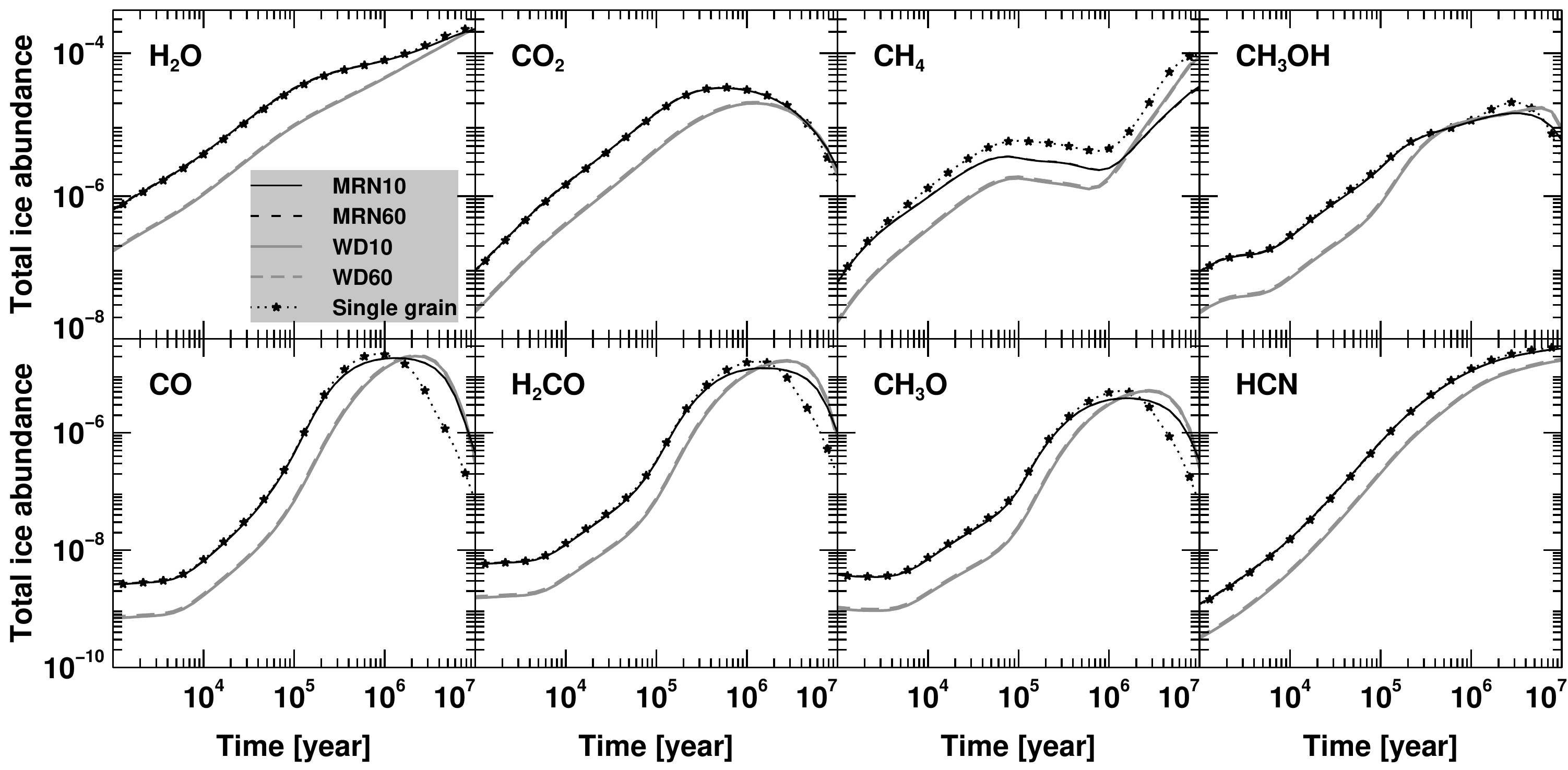}
   \caption{The total ice abundances of selected species on dust grains as a function of time for models with different grain size distributions and a different number of bins. Legends apply to all panels.}
   \label{Fig:5}
\end{figure*}
\begin{figure*}
    \centering
   \includegraphics[width=.97\textwidth,trim = 0cm 0cm 0cm 0cm, clip,angle=0]{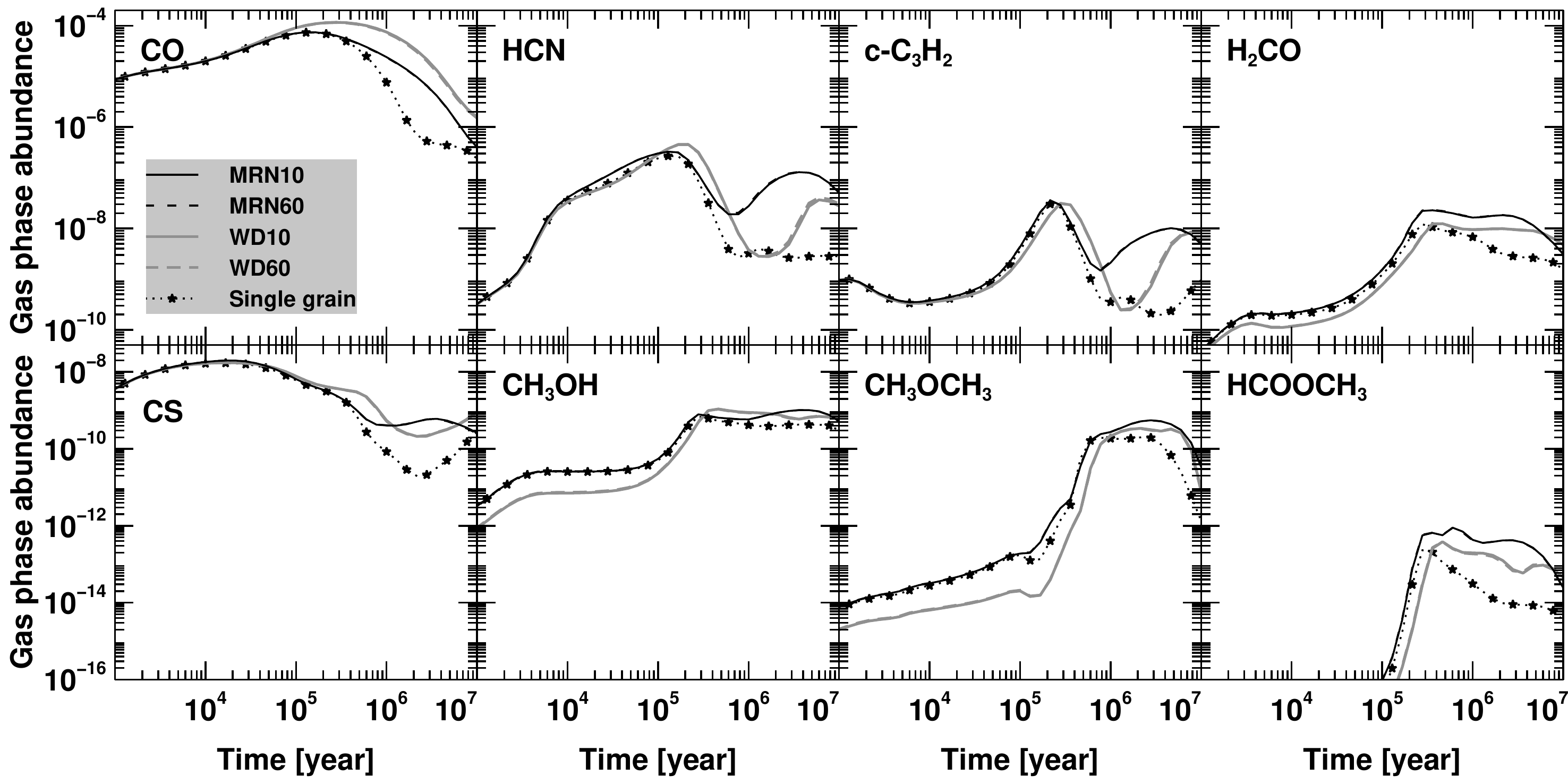}
   \caption{Gas phase abundances of selected species as a function of time for models with different grain size distributions and a different number of bins. Legends apply to all panels.}
   \label{Fig:7_2}
\end{figure*}

In this section, we explore the effect of varying the number of grain size bins in our model (1, 10, 30 or 60) and analyze how models with different grain size distribution (the MRN and the WD) compare to each other. We keep the surface temperature at 10K for all grain sizes and the cosmic-ray-induced desorption is also enabled in all models.

Figure \ref{Fig:3} shows the normalized total abundance of species on grain surface as a function of time. Here we have summed up the abundances of all species on all grain sizes in the multi-grain model to compare with the total ice abundance in the single-grain model. Further, to better visualize the effect of grain size distribution,   we normalized the total ice abundance in all models with the total ice abundance obtained in the single grain model.  In both the MRN and the WD grain size distributions, we observe slight differences in the results due to a difference in the number of grains used in simulations. In both models, the total abundance using 30 bins and 60 bins is almost the same and slightly higher than when using 10 bins (black solid line for the MRN model and gray solid line for the WD model). 

  We see in Fig. \ref{Fig:3} that initially (time $< 10^3$ yr) the model MRN60, which has the highest total effective surface area (see Table \ref{tbl:T-eff_area}), has maximum ice. A higher surface area results in more accretion of gaseous species. At this early stage, the ice abundance is very low so the effect of cosmic-ray-induced desorption is not visible. But as ice abundance increases the effect of cosmic ray heating increases and we notice that relative abundance of ice in the MRN case becomes lower than that in the single-grain model and remains so until the end of the simulation. However, this relative change in abundance is not unidirectional. We see that the relative abundance in the MRN case starts increasing around $8\times10^5$ yr and then again at around $7\times10^6$ yr. But this trend is not sustained for long. This fluctuation in the relative abundance may be because of a shortage of volatile species on the surface resulting in reduced desorption via the cosmic ray hitting until the grains are populated again with the volatile species.   

In the model with the WD grain size distribution, initial ice abundance is very low (about 25\% only) compared to the other two models  but it increases with time and comes very close to the total ice abundance in the single-grain model by the end of the simulation. The WD grain size distribution is very different from the MRN grain size distribution in two major ways. First, the total number density of grains is smaller by a large factor resulting in a lower total effective surface area by a factor of almost 3.5. The effect of this is a lower ice production rate under similar gas phase due to a lower accretion rate, so initial ice abundance is very low.  Second, the total effective surface area of the   small  grains in the WD distribution is less than that of the   big  grains (see top panel in Fig. \ref{Fig:grainAbn}) while in the MRN distribution it is opposite. In other words, more ice is formed on the   big  grains than the   small  grains. The obvious effect of this is a reduced effect of the cosmic-ray collisions as ices on the   small  grains are affected the most by the cosmic ray collisions. We have seen that the gas phase abundance of N-bearing species is strongly affected by the cosmic-ray collisions with the dust grains. 

In Fig. \ref{Fig:N_species_WD} we plot the gas phase abundances of a few N-bearing species for the WD case. We see that the differences in the results for models with and without the cosmic-ray-induced desorption are smaller than in the single grain model (Fig. \ref{Fig:N_species}). Specifically, the abundance of N$_2$ and N is not affected as strongly as seen earlier. After $3\times10^6$ yr, the effect of the cosmic-ray-induced desorption starts to strongly affect the chemistry. This is expected as by this time there are lots of ices on the   small  grains which can desorb due to the cosmic-ray collisions. 

Figure \ref{Fig:5} shows the total ice abundances of selected species on the dust grains as a function of time for different grain size distributions and different bin numbers. In each multi-grain model, we summed the surface abundances of selected species over all grain sizes to get the total abundance of that species at any specific moment in time. The results of the model with the MRN distribution were discussed in great detail in the previous section. Here we note that the total ice abundances of H$_2$O, CO$_2$ and HCN in the MRN model are very similar to the single-grain model and higher than that in the WD model. In the previous section we saw that the ice compositions on different grain sizes are very different, with the   small  grains having a higher percentage of heavier species and a lower percentage of lighter species, but, conversely, here we see that the total ice abundances of some species may remain similar in both the MRN model and the single grain model. The main difference between the models with the two different size distributions seems to be a shift in the time. In the model with the WD distribution, having a smaller total dust density and the surface area, the ices grow very slowly but at a very steady rate as depletion of gas phase species (and so the efficiency of the surface reactions) occurs at a much lower rate compared to the other two models.   In the WD case, at the end of the simulation (10$^7$ yr), the ice abundances of most of the species are more than that obtained in the MRN case. This is due to less efficient desorption via cosmic ray collisions in the WD model. This helps in keeping most of the accreted matter on the grain surface.  

When looking at the gas phase abundances (see Fig. \ref{Fig:7_2}), we see that many species are more abundant in the WD case than in the simulation with the MRN distribution or the single grain size model. For the WD model, the species formed in the gas phase only, such as CO, CS, and O$_2$, present a larger gas-phase abundances as there are less grains to deplete them. On the contrary, species formed on the grains (such as CH$_3$OH, HCOOCH$_3$ or CH$_3$OCH$_3$) have smaller abundances as there are less grains to produce them. But abundances of these species start to increase significantly after 10$^5$ yr and by the time the simulation reaches $10^6$ yr, most of these species have gas phase abundances much higher than that in the single grain model. This rapid gain in the gas phase abundances of certain species is due to the desorption from the   small  grains. For example, at $3\times10^5$ yr, desorption of CH$_3$OCH$_3$ ice from the smallest grain contributes almost 5\% of the production of CH$_3$OCH$_3$ in the gas phase while the contribution of the second smallest grain is about 1.5\%. This contribution is absent in the single-grain model. Simulations with the MRN distribution always give higher gas phase abundances for most of these species.

As a general result, the choice of grain size distribution can have a strong impact on the gas-phase abundances of some species. Simulation with the WD distribution shows that the   small  grains in the simulation can play a vital role at later times even if their number density is lower. Here, we also note that although using 10 grain sizes changes the results significantly over the single grain model irrespective of which type of grain size distribution is used, changing to 30 or 60 grain sizes has little impact. Although implementation of more processes which are grain-size-dependent, such as fluctuation in the surface temperature due to UV photons hitting the grains or nonuniform surface temperature, may result in very significant changes with more grain sizes.

\subsection{Effect of dust-size-dependent surface temperature}
\label{R_sec:3}
\begin{figure}
    \centering
   \includegraphics[width=.47\textwidth,trim = 0cm 0cm 0cm 0cm, clip,angle=0]{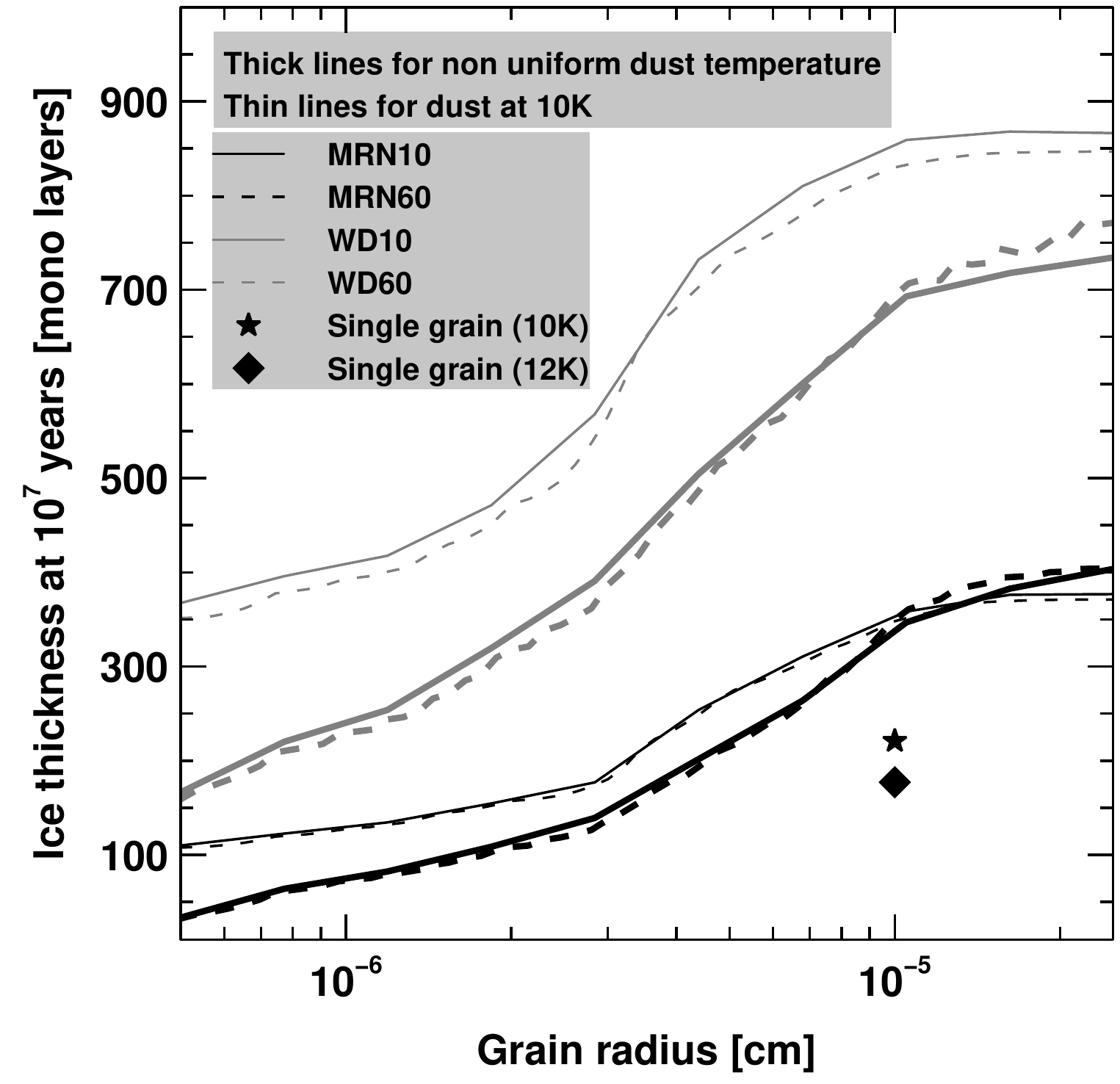}
   \caption{The ice thickness as a function of grain radius at $10^7$ yr in the different models. Thin lines: all grains have the same temperature of 10K. Thick lines: all grains have their own temperature, see Fig. \ref{Fig:grainTemp}. Cosmic ray induced desorption is used in all models.}
   \label{Fig:4}
\end{figure}
\begin{figure*}
    \centering
   \includegraphics[width=.97\textwidth,trim = 0cm 0cm 0cm 0cm, clip,angle=0]{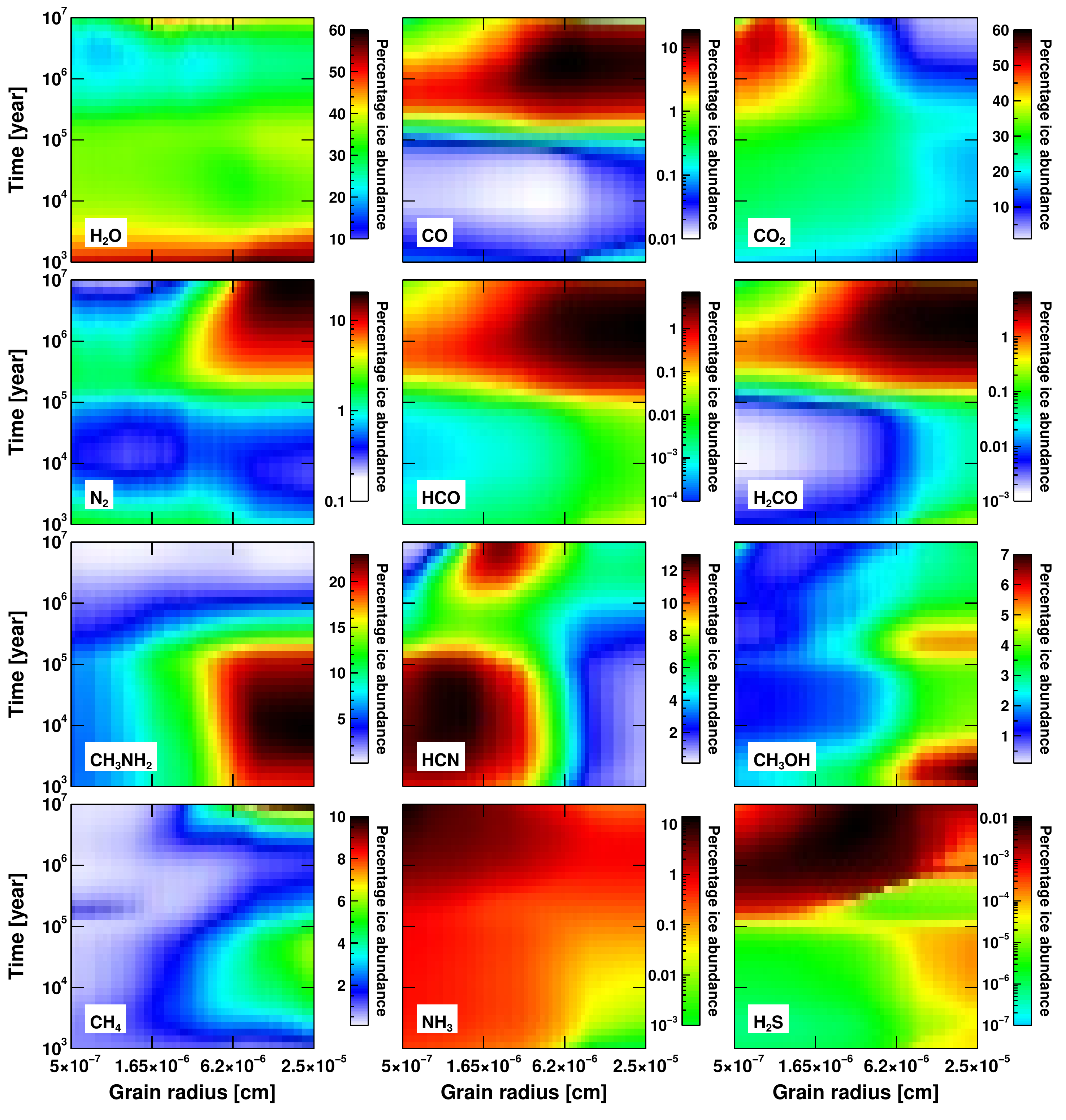}
   \caption{Percentage ice abundances of the most abundant species as a function of grain size  (x-axis) and time (y-axis). Results are for the multi grain model with 60 grain sizes and with non-uniform surface temperatures. The cosmic-ray-induced desorption is used and the MRN grain size distribution was used to calculate number density of 60 grain bins.}
   \label{Fig:Per_ab_map_VT}
\end{figure*}
\begin{figure*}
    \centering
   \includegraphics[width=.97\textwidth,trim = 0cm 0cm 0cm 0cm, clip,angle=0]{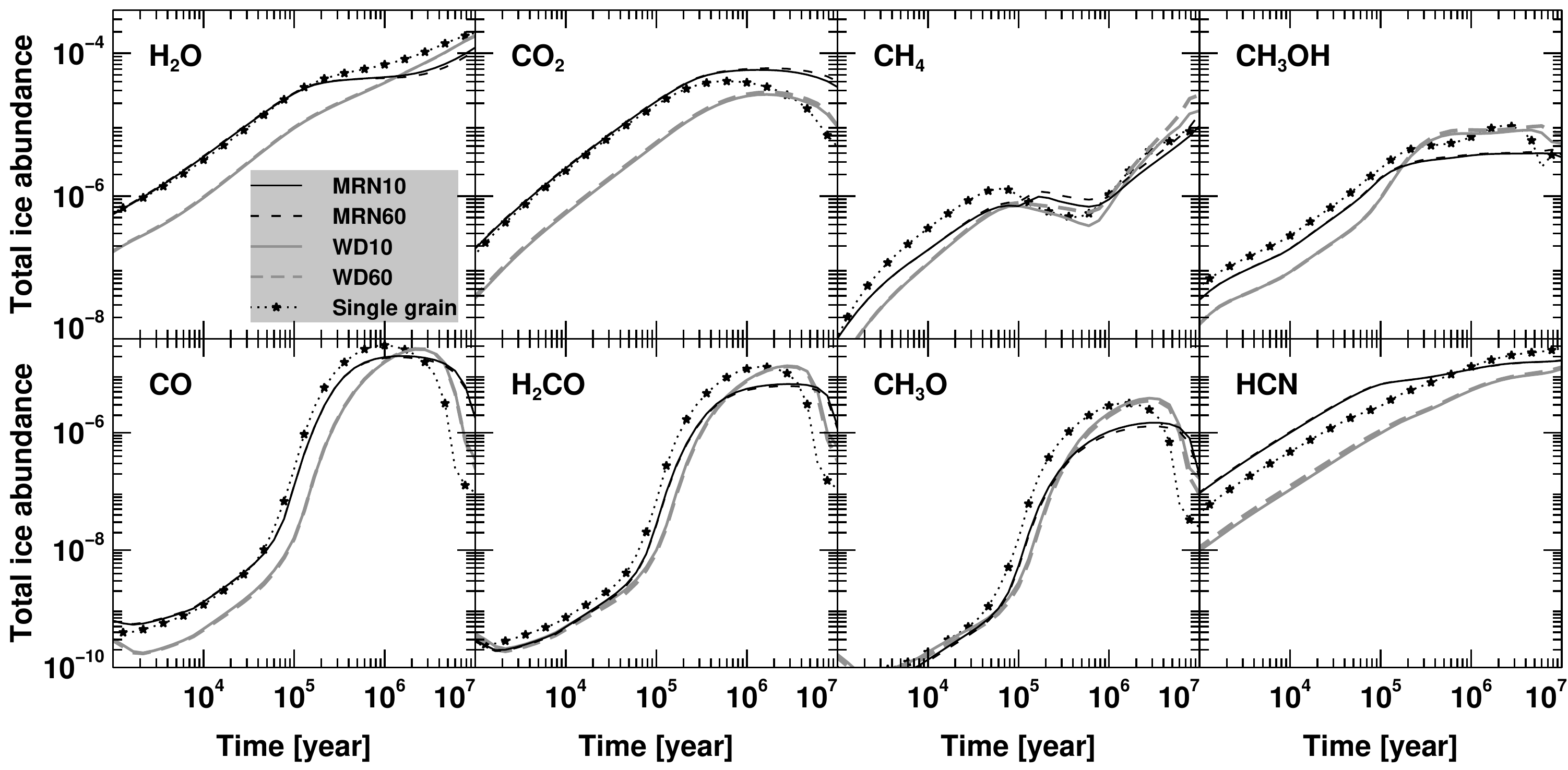}
   \caption{Total ice abundances of selected species on dust grains as a function of time for the models with different grain size distributions and different number of bins. Grain temperature is non-uniform in the multi grain models and in the single grain model surface temperature is 12K. Legends apply to all panels.}
   \label{Fig:5c}
\end{figure*}
\begin{figure*}
    \centering
   \includegraphics[width=.97\textwidth,trim = 0cm 0cm 0cm 0cm, clip,angle=0]{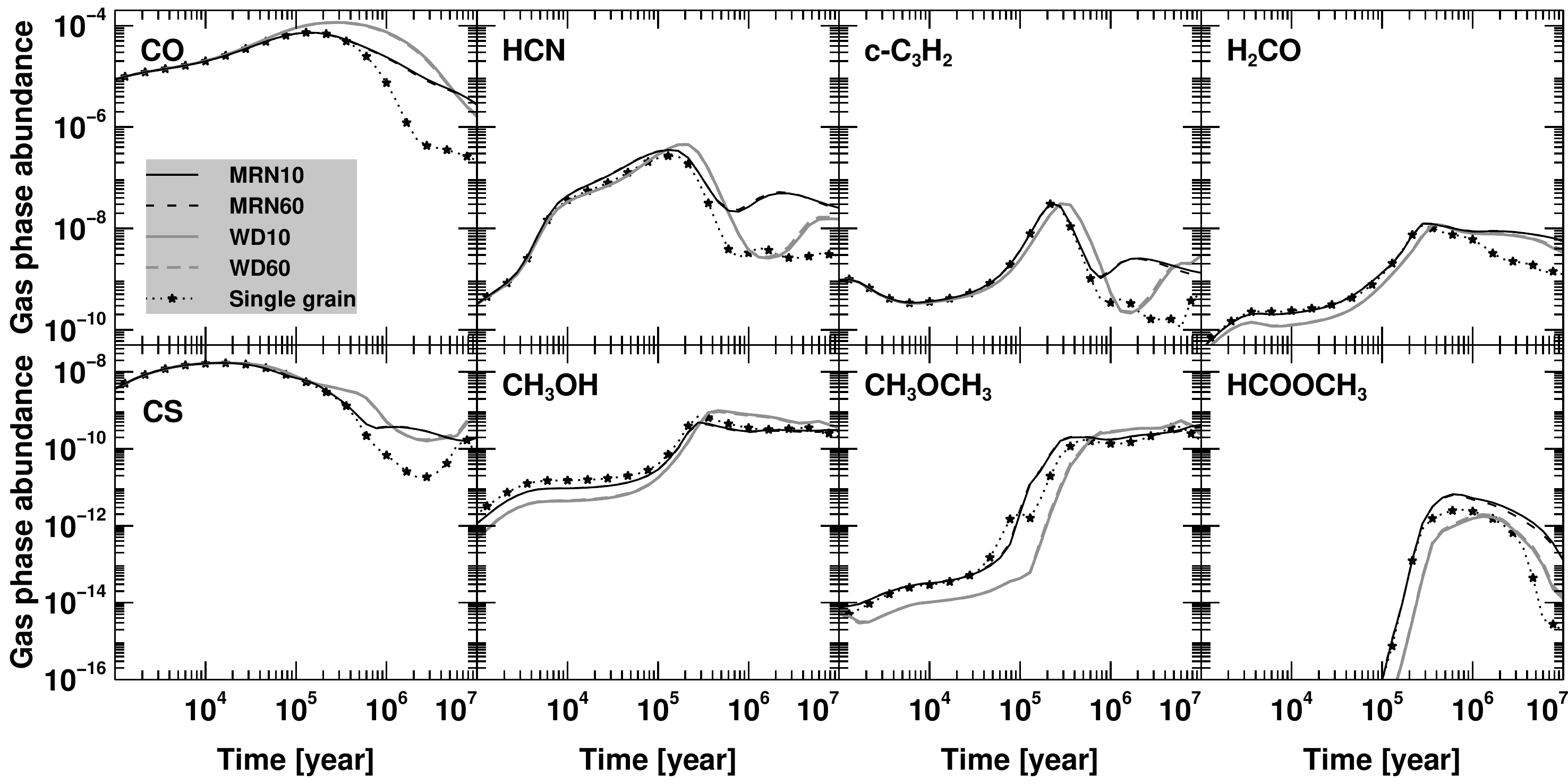}
   \caption{Gas phase abundances of selected species as a function of time for models with different grain size distributions and a different number of bins. Grain temperature is nonuniform in the multi-grain models and in the single-grain model surface temperature is 12K. This legend applies to all panels.}
   \label{Fig:7c}
\end{figure*}
In the diffuse clouds, the surface temperature varies strongly with the grain size (\cite{Draine1984}) but in the dense cold clouds it is also well accepted that grains of different sizes have different temperatures. The surface temperature is an important parameter and can change the ice composition significantly. In this section, we explore the effect of considering a dust temperature that would depend on the size of the grains. The bottom panel in Fig. \ref{Fig:grainTemp} shows the surface temperatures of the grains as a function of their radii used in the models. The grain temperature varies between 13.7K (for the smallest $0.005\mu$m grains) and 11.1K (for the biggest $0.25\mu$m grains). We note that in this model, the grain temperature of the single grain size model is 12K (see Sect. \ref{M_sec:1}). A modification of the surface temperature effectively changes all the reaction rates on the surface but the diffusion of species with the smaller diffusion energies are the most affected because the diffusion rate increases exponentially with the surface temperature. Also, a change in the temperature affects mostly the chemistry in the top two layers as reactivity is small in the mantles.  

Figure \ref{Fig:4} shows the ice thickness at the end of simulations ($10^7$ yr) as a function of grain radius for different models: the single-grain model (symbols) and the multi-grain models with 10 (dashed lines) and 60 (solid lines) grain sizes (both the WD and the MRN distributions are shown). Here, we show two different cases: 1) All grains are at the same temperature of 10K (thin lines for the multi-grain model and star for the single-grain model), and 2) all grains have their own temperature (thick lines for the multi-grain model and a diamond for the single grain model). The ice thickness in the simulations with the WD distribution is more than twice that in other models. This is because a smaller number density of grains results in more accretion on each grain size.

In the case of the size-dependent temperature model, the   big  grains have a temperature of about 11K, which is a small difference as compared to the model with grains at 10K. This small difference has however a strong impact on the ice thickness for the WD model resulting in a difference of about 100 monolayers (i.e., 11\% of the total ice thickness). This difference is even larger for the   small  grains. As the   small  grains have an even higher temperature than the   big grains, the difference for the small grains between fixed and size-dependent temperature is larger. The effect is also seen for the MNR model in that case. 

We now compare the ice thickness for models with different numbers of bins (10 and 60). We see that in both (the MRN and the WD) multi-grain models with all grains at 10K, the model with 10 bins has slightly higher ice thickness as compared to the model with 60 bins. But in the case of nonuniform surface temperature,   big  grains (radii $> 0.09\mu$m) in the model with 60 bins have more ices compared to   big  grains in the model with 10 grains. We assume that this is due to the difference in the bin sizes used in these models. We know that in the model with 60 bins, each bin is very small compared to the model with 10 bins, specially bins for   big  grains due to logarithmic division (see Fig. \ref{Fig:grainBin}). Keeping the bin size small helps in better approximation of the surface temperature as grains in a small bin can have temperatures that are very similar, but if the bin size is big, it may not be valid to assume that all grains in that bin are at the same temperature.

In the case of the WD model, the ice thickness is always smaller in the case of nonuniform temperature as compared to the 10K model. For the MNR case, on the contrary, the mass transfer makes the curves cross for grains around $0.1\mu$m. For the big grains, the ice thickness is larger with a nonuniform temperature as compared to the 10K model.

Figure \ref{Fig:Per_ab_map_VT} shows colored maps of percentage ice abundances of selected species as a function of grain size and time in years for the model using the MRN distribution. We compared this figure with Fig. \ref{Fig:Per_ab_map} in which the surface temperature is constant at 10K for all grain sizes. We notice that the percentage of water on the   small  grains has reduced significantly. In this model, CO$_2$ ice is the most abundant on the   small  grains, while in the model with a uniform surface temperature  H$_2$O is the most abundant. The percentage of CO$_2$ ice has increased by more than a factor of two. We know that the obvious effect of higher surface temperature is an increase of hopping and desorption rates. In the case of CO$_2$, this is essentially due to increased mobility of O that results in more efficient production of CO$_2$. Opposite to this, a higher surface temperature increases the desorption rate of H$_2$ so significantly that water production goes down sharply. In general we see that, the percentage ice abundances of all species change with grain radii and this change in the ice compositions are stronger than in the uniform surface temperature case. To see the effect of this on the total ice productions of various species, we plot, in Fig. \ref{Fig:5c},  the total abundances of selected species on the dust grains as a function of time for different models with grain-size-dependent temperature. Here we can clearly see that, in the MRN case, the total ice abundance of water has gone down compared to the model with uniform surface temperature while the same for CO$_2$ has gone up (see Fig. \ref{Fig:5} for comparison). The total abundance of water ice is almost unchanged in the single-grain model and the WD distribution, while the CO$_2$ ice abundance has increased but not very significantly. This is because the   small  grains play a major role in determining ice composition in the MRN model. The total ice abundance of most of the species has decreased due to the higher surface temperature. We also notice that, for  a
few species, results with 10 and 60 grain sizes do not overlap, which was the case in the previous results for the uniform surface temperature. Differences are especially visible for CH$_4$ and CH$_3$OH.

Figure \ref{Fig:7c} shows the gas phase abundances for selected species as a function of time for different models with grain-size-dependent temperature. We see very similar effects in the gas phase. The abundance of CO has increased in the MRN case as compared to Fig. \ref{Fig:7_2} (uniform 10K grain temperature) while it is almost unchanged for the WD models. Abundance of CH$_3$OH has decreased as it is now produced less efficiently on grains due to the higher desorption rate of adsorbed H atoms. Species that are mainly formed in gas phase, such as H$_2$CO or HCOOCH$_3$, have higher abundances.

\section{Comparison with observations in cold dense clouds}
\label{sec:comp}
\begin{figure*}
    \centering
   \includegraphics[width=.97\textwidth,trim = 0cm 0cm 0cm 0cm, clip,angle=0]{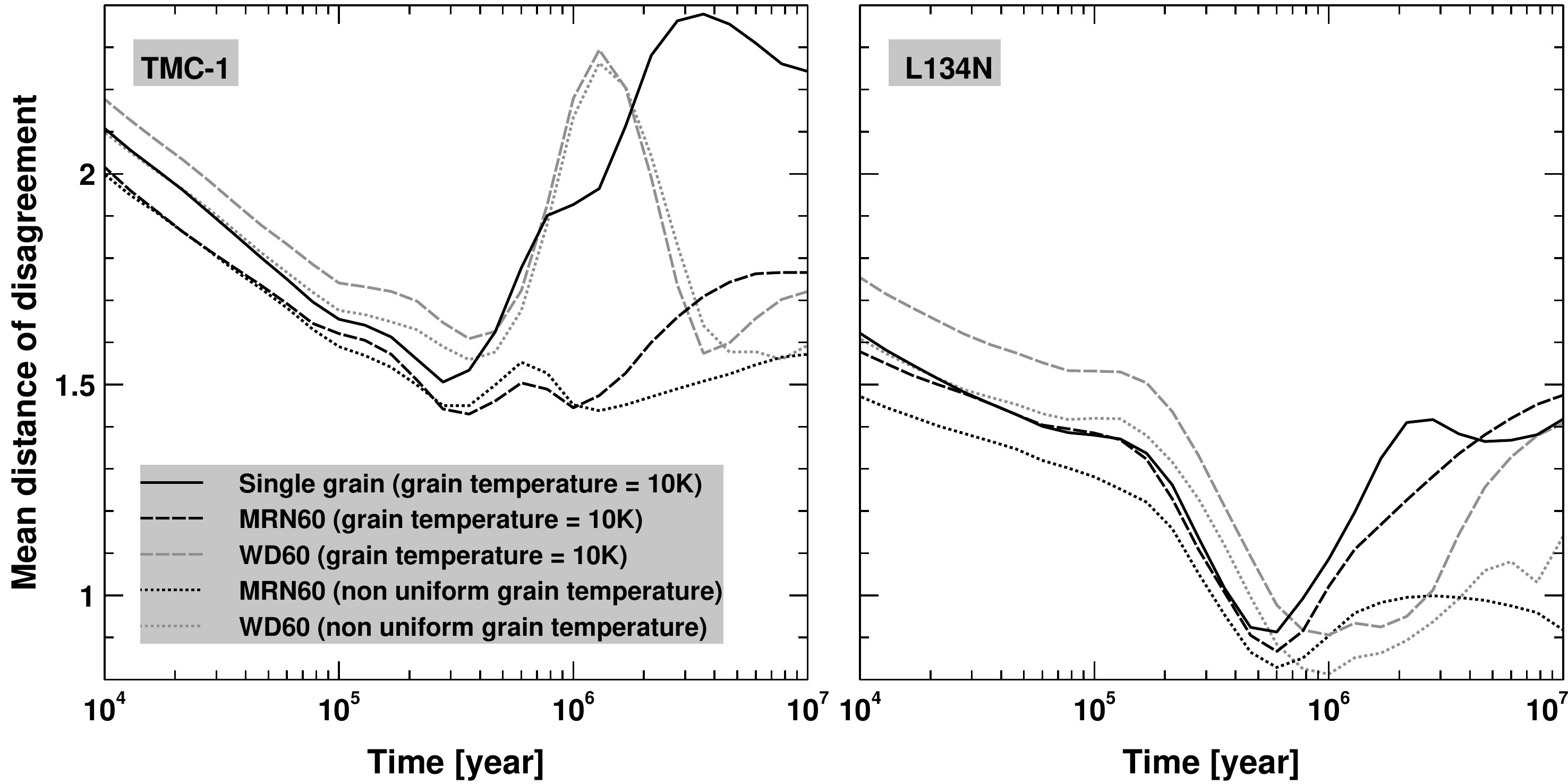}
   \caption{Mean distance of disagreement computed for two dark clouds TMC-1 network and L134N.
}
   \label{Fig:MDDA}
\end{figure*}
We compared our results obtained from different models with observations in the cold dense clouds TMC-1 and L134N. For comparison, we used the
method described in \cite{Loison2014}. We have computed the mean distance of disagreement $D(t)$ for each output at time $t$ of the simulation using the following formula
\begin{equation}
 D(t) = \frac{1}{N_{obs}} \sum_{i}|[log[n(X_i,t)]-log[n(X_i^{obs})]]|
,\end{equation}
where $n(X_i,t)$ is the calculated abundance of species $X_i$  at time $t$ , $n(X_i^{obs})$  is the observed abundance for the same species, and $N_{obs}$ is the total number of observed species used in the calculation of $D(t)$. In this method, the smaller the value of $D(t)$, the better the agreement between the observed abundances and the simulated results. We used tabulated data in \cite{Agundez2013} for observation data. From the table we used abundances of 34 molecules for L134N and 52 molecules for TMC-1; we did not consider species for which only upper or lower limits are given. 

Figure \ref{Fig:MDDA} shows the calculated mean distance of disagreement from the observed abundances in the two clouds as a function of time as obtained from the single-grain and multi-grain models. First we will discuss the case of TMC-1. We see that all models give the best agreement with the observation at about $2\times10^5$ yr. At this time, the multi-grain model with the MRN distribution (black dashed and dotted lines) gives the best agreement followed by the single grain model (black solid line). After reaching this point of best agreement, both the single-grain model and the multi-grain model with the WD distribution deviate away and cross the line where the mean distance of disagreement becomes more than two, while the multi-grain model with the MRN distribution remains best. The multi-grain model with the WD distribution again returns to the best agreement at about $2\times10^6$ yr. In the multi-grain models, we see that the model with nonuniform surface temperatures (dotted lines) gives the best agreement for both the MRN and the WD distributions. The reason why the multi-grain models stay close to the best value while the single-grain model does not is again due to basic differences in the two models. The single-grain model  lacks a proper desorption mechanism for the ices. Due to this there is no way to cycle material from the gas to the grain and then back to the gas. This results in a gradual increase in the ice abundance only. So after reaching a state of the best agreement there is no way that it can fluctuate around it. In the multi-grain model, due to the presence of a large number of   small  grains, we have a better desorption mechanism in the form of the cosmic-ray-induced desorption that results in a cycle of material from the grains to the gas. 

We see a very similar result in the case of L134N. All models give a very good value of $D(t)$ at around $6\times10^5$ yr. At this time $D(t)$ is less than one for all models. This means that the mean difference between modeled and observed abundances is smaller than a factor of ten. This case also shows that the multi-grain model with the WD distribution gives better agreement than the same model with the MRN distribution. This essentially shows how different types of grain size distribution can lead to differing estimates of the observed abundance in different clouds.

\section{Discussion on the cooling of grains after a collision with cosmic ray particles}
\label{sec:discussion}

The cooling of the grains after the collisions with cosmic-ray particles is assumed to be the most efficient by the evaporation of the surface species. The cooling time of $10^{-5}$ s estimated by \cite{Hasegawa1993} \citep[see also Sect. \ref{M_sec:1} and][]{Herbst2006} is based on the assumption that the grains are covered by a significant number of CO molecules. We have shown that the   small  grains are depleted in light species such as CO due to the higher temperatures with respect to the larger grains. Since other abundant ice species (H$_2$O, N$_2$, CH$_4$, CO$_2$, or CH$_3$OH) have higher desorption energies, the cooling of the smallest grains may take longer than what is assumed here. 

To test the sensitivity of the model to this, we have scaled the duration of the temperature peaks produced by the cosmic ray collisions for grains smaller than $0.1\mu$m. In fact the   small  grains have less sites so less possibility for CO to desorb. For the scaling, we used a simple factor given by $10^6/N_s(i)$, where $N_s(i)$ is the number of sites on the $i$th grain.  Using this scaling, the desorption from the   small  grains is more efficient. Looking at the complex organic molecules observed in the cold cores, this additional nonthermal desorption may increase their gas phase abundances. All of these   complex organic molecules (COMs)  are, however, not necessarily produced enough on the grains, or, rather, are  more abundant on the big grains than on the small ones. Here we do not take into account the composition of the grains themselves (i.e., the presence of CO). 

In our model, methanol is very abundant on small grains and more desorption from the small grains could account for the observed gas phase abundances \citep{Oberg2010}. Similar results were found for CH$_3$O, CH$_3$CCH, HCOOH, and CH$_3$OCH$_3$. But while CH$_3$CHO is mostly present on   big  grains ($>0.02\mu$m), the ice abundances of HCOOCH$_3$ on grains (whatever the size) are too low to account for the observed gas phase abundances.

\section{Conclusions}
Here we summarize our main findings from this work.

We observe very small differences in the results when switching from 10 grain sizes to 30 grain sizes in our models and almost no difference in the results for simulations with 30 grain sizes and 60 grain sizes. Therefore, for both the MRN and the WD models, one can use up to 30 grain sizes if better precision is needed and there is almost no need to go above 30 grain sizes. This result may change if one tries with some other grain size distribution model or stochastic heating of dust grains.

Results from the single-grain and multi-grain models both using the MRN distribution are very close and are similar in absence of the cosmic rays provided all grains are at the same surface temperature. This is due to a very similar total effective surface area in both the models.

The collisions of the cosmic ray particles with the   small  grains ($<0.04\mu$m) causes the evaporation of a large portion of the ices from the surfaces. This results in a significant reduction in the ice thickness on these grains. The evaporated species then accrete again on the surfaces of the big grains, on which the cosmic ray induced peak temperature is much smaller. The cosmic-ray-induced desorption thus produces a mass transfer of ices from   smaller   to   bigger  grains. The mass transfer is even more efficient when the size dependency of the dust temperature is considered. 

The choice of the MRN or the WD distributions strongly affects the abundances of most of the species. The cosmic-ray-induced desorption is the most effective in the MRN distribution case. In the WD distribution case, a larger contribution to the total effective surface area comes from   big  grains, and therefore most of the ices are formed on the   big  grains.    Also, for most species, the cosmic-ray-induced desorption of the ices brings about noticeable changes in the gas phase abundances after $10^6$ yr only  when there are significant ice abundances on the   small  grains. Most of the species have smaller ice abundances in the WD case due to a lower number density of dust grains as compared to the MRN distribution. 

Ice composition is dependent on grain size. Ices on the   small  grains contain a   small  percentage of lighter molecules, such as CO, HCO and N$_2,$ and a larger percentages of more strongly bound species, such as water and CO$_2$. 

Considering a nonuniform surface temperature for the different grain sizes strongly impacts the overall gas and ice compositions. The difference in the chemical compositions between the   small  and   big  grains is even stronger. 

The MRN distribution gives a better agreement with TMC-1 observations whether the dust temperature is uniform or not, while both the MNR and the WD distributions give a better agreement with L134N when dust temperature is nonuniform. 

\section{Acknowledgements}
This study has received financial support from the French State  in  the  frame  of  the  "Investments  for  the  future"  Programme  IdEx  Bordeaux, reference ANR-10-IDEX-03-02. VW research is funded by an ERC Starting Grant (3DICE, grant agreement 336474) and the CNRS program Physique et Chimie du Milieu Interstellaire (PCMI) co-funded by the Centre National d'Etudes Spatiales (CNES). V.W. thanks Bruce Draine for helpful discussions.
%
%
%
%
%
%
%
%
\bibliography{aa32486} 

\begin{thebibliography}{}

\bibitem[{Acharyya} et~al., 2011]{Acharyya2011}
{Acharyya}, K., {Hassel}, G.~E., and {Herbst}, E. (2011).
\newblock {The Effects of Grain Size and Grain Growth on the Chemical Evolution
  of Cold Dense Clouds}.
\newblock {\em \apj}, 732:73.

\bibitem[{Ag{\'u}ndez} and {Wakelam}, 2013]{Agundez2013}
{Ag{\'u}ndez}, M. and {Wakelam}, V. (2013).
\newblock {Chemistry of Dark Clouds: Databases, Networks, and Models}.
\newblock {\em Chemical Reviews}, 113:8710--8737.

\bibitem[{Draine} and {Lee}, 1984]{Draine1984}
{Draine}, B.~T. and {Lee}, H.~M. (1984).
\newblock {Optical properties of interstellar graphite and silicate grains}.
\newblock {\em \apj}, 285:89--108.

\bibitem[{Garrod}, 2008]{Garrod2008}
{Garrod}, R.~T. (2008).
\newblock {A new modified-rate approach for gas-grain chemical simulations}.
\newblock {\em \aap}, 491:239--251.

\bibitem[{Ge} et~al., 2016]{Ge2016}
{Ge}, J.~X., {He}, J.~H., and {Li}, A. (2016).
\newblock {Interstellar chemical differentiation across grain sizes}.
\newblock {\em \mnras}, 460:L50--L54.

\bibitem[{Graedel} et~al., 1982]{Graedel82}
{Graedel}, T.~E., {Langer}, W.~D., and {Frerking}, M.~A. (1982).
\newblock {The kinetic chemistry of dense interstellar clouds}.
\newblock {\em \apjs}, 48:321--368.

\bibitem[{Hasegawa} and {Herbst}, 1993a]{Hasegawa1993}
{Hasegawa}, T.~I. and {Herbst}, E. (1993a).
\newblock {New gas-grain chemical models of quiescent dense interstellar clouds
  - The effects of H2 tunnelling reactions and cosmic ray induced desorption}.
\newblock {\em \mnras}, 261:83--102.

\bibitem[{Hasegawa} and {Herbst}, 1993b]{Hasegawa1993b}
{Hasegawa}, T.~I. and {Herbst}, E. (1993b).
\newblock {Three-Phase Chemical Models of Dense Interstellar Clouds - Gas Dust
  Particle Mantles and Dust Particle Surfaces}.
\newblock {\em \mnras}, 263:589.

\bibitem[{Hasegawa} et~al., 1992]{Hasegawa1992}
{Hasegawa}, T.~I., {Herbst}, E., and {Leung}, C.~M. (1992).
\newblock {Models of gas-grain chemistry in dense interstellar clouds with
  complex organic molecules}.
\newblock {\em \apjs}, 82:167--195.

\bibitem[{Herbst} and {Cuppen}, 2006]{Herbst2006}
{Herbst}, E. and {Cuppen}, H.~M. (2006).
\newblock {Interstellar Chemistry Special Feature: Monte Carlo studies of
  surface chemistry and nonthermal desorption involving interstellar grains}.
\newblock {\em Proceedings of the National Academy of Science},
  103:12257--12262.

\bibitem[{Hincelin} et~al., 2011]{Hincelin2011}
{Hincelin}, U., {Wakelam}, V., {Hersant}, F., {Guilloteau}, S., {Loison},
  J.~C., {Honvault}, P., and {Troe}, J. (2011).
\newblock {Oxygen depletion in dense molecular clouds: a clue to a low O$_{2}$
  abundance?}
\newblock {\em \aap}, 530:A61.

\bibitem[{Iqbal} et~al., 2014]{iqbal2014}
{Iqbal}, W., {Acharyya}, K., and {Herbst}, E. (2014).
\newblock {H$_{2}$ Formation in Diffuse Clouds: A New Kinetic Monte Carlo
  Study}.
\newblock {\em \apj}, 784:139.

\bibitem[{Jenkins}, 2009]{Jenkins09}
{Jenkins}, E.~B. (2009).
\newblock {A Unified Representation of Gas-Phase Element Depletions in the
  Interstellar Medium}.
\newblock {\em \apj}, 700:1299--1348.

\bibitem[{L\'{e}ger} et~al., 1985]{Leger1985}
{L\'{e}ger}, A., {Jura}, M., and {Omont}, A. (1985).
\newblock {Desorption from interstellar grains}.
\newblock {\em \aap}, 144:147--160.

\bibitem[{Loison} et~al., 2014]{Loison2014}
{Loison}, J.-C., {Wakelam}, V., {Hickson}, K.~M., {Bergeat}, A., and {Mereau},
  R. (2014).
\newblock {The gas-phase chemistry of carbon chains in dark cloud chemical
  models}.
\newblock {\em \mnras}, 437:930--945.

\bibitem[{Mathis} et~al., 1983]{Mathis1983}
{Mathis}, J.~S., {Mezger}, P.~G., and {Panagia}, N. (1983).
\newblock {Interstellar radiation field and dust temperatures in the diffuse
  interstellar matter and in giant molecular clouds}.
\newblock {\em \aap}, 128:212--229.

\bibitem[{Mathis} et~al., 1977]{Mathis1977}
{Mathis}, J.~S., {Rumpl}, W., and {Nordsieck}, K.~H. (1977).
\newblock {The size distribution of interstellar grains}.
\newblock {\em \apj}, 217:425--433.

\bibitem[{{\"O}berg} et~al., 2010]{Oberg2010}
{{\"O}berg}, K.~I., {Bottinelli}, S., {J{\o}rgensen}, J.~K., and {van
  Dishoeck}, E.~F. (2010).
\newblock {A Cold Complex Chemistry Toward the Low-mass Protostar B1-b:
  Evidence for Complex Molecule Production in Ices}.
\newblock {\em \apj}, 716:825--834.

\bibitem[{Pauly} and {Garrod}, 2016]{Pauly2016}
{Pauly}, T. and {Garrod}, R.~T. (2016).
\newblock {The Effects of Grain Size and Temperature Distributions on the
  Formation of Interstellar Ice Mantles}.
\newblock {\em \apj}, 817:146.

\bibitem[{Ruaud} et~al., 2016]{Ruaud2016}
{Ruaud}, M., {Wakelam}, V., and {Hersant}, F. (2016).
\newblock {Gas and grain chemical composition in cold cores as predicted by the
  Nautilus three-phase model}.
\newblock {\em \mnras}, 459:3756--3767.

\bibitem[{Umebayashi} and {Nakano}, 1981]{Umebayashi1981}
{Umebayashi}, T. and {Nakano}, T. (1981).
\newblock {Fluxes of Energetic Particles and the Ionization Rate in Very Dense
  Interstellar Clouds}.
\newblock {\em \pasj}, 33:617.

\bibitem[{Wakelam} et~al., 2013]{Wakelam2013}
{Wakelam}, V., {Cuppen}, H.~M., and {Herbst}, E. (2013).
\newblock {\em {Astrochemistry: Synthesis and Modelling}}, pages 115--143.
\newblock Springer Berlin Heidelberg, Berlin, Heidelberg.

\bibitem[{Wakelam} and {Herbst}, 2008]{Wakelam08}
{Wakelam}, V. and {Herbst}, E. (2008).
\newblock {Polycyclic Aromatic Hydrocarbons in Dense Cloud Chemistry}.
\newblock {\em \apj}, 680:371--383.

\bibitem[{Wakelam} et~al., 2015]{Wakelam2015}
{Wakelam}, V., {Loison}, J.-C., {Herbst}, E., {Pavone}, B., {Bergeat}, A.,
  {B{\'e}roff}, K., {Chabot}, M., {Faure}, A., {Galli}, D., {Geppert}, W.~D.,
  {Gerlich}, D., {Gratier}, P., {Harada}, N., {Hickson}, K.~M., {Honvault}, P.,
  {Klippenstein}, S.~J., {Le Picard}, S.~D., {Nyman}, G., {Ruaud}, M.,
  {Schlemmer}, S., {Sims}, I.~R., {Talbi}, D., {Tennyson}, J., and {Wester}, R.
  (2015).
\newblock {The 2014 KIDA Network for Interstellar Chemistry}.
\newblock {\em \apjs}, 217:20.

\bibitem[{Wakelam} et~al., 2010]{Wakelam2010}
{Wakelam}, V., {Smith}, I.~W.~M., {Herbst}, E., {Troe}, J., {Geppert}, W.,
  {Linnartz}, H., {{\"O}berg}, K., {Roueff}, E., {Ag{\'u}ndez}, M., {Pernot},
  P., {Cuppen}, H.~M., {Loison}, J.~C., and {Talbi}, D. (2010).
\newblock {Reaction Networks for Interstellar Chemical Modelling: Improvements
  and Challenges}.
\newblock {\em \ssr}, 156:13--72.

\bibitem[{Weingartner} and {Draine}, 2001]{Weingartner2001}
{Weingartner}, J.~C. and {Draine}, B.~T. (2001).
\newblock {Dust Grain-Size Distributions and Extinction in the Milky Way, Large
  Magellanic Cloud, and Small Magellanic Cloud}.
\newblock {\em \apj}, 548:296--309.

\end{thebibliography}

\end{document}